\PassOptionsToPackage{table}{xcolor}
\documentclass[journal]{IEEEtran}
\ifCLASSINFOpdf
\else
\fi

\usepackage[utf8]{inputenc}
\usepackage{amsmath}
\usepackage{mathtools}
\usepackage{float}
\usepackage{braket}
\usepackage{graphicx}
\graphicspath{{./figures/}}
\usepackage{amsfonts}
\usepackage[ruled,vlined,linesnumbered]{algorithm2e}
\let\oldnl\nl% Store \nl in \oldnl
\newcommand{\nonl}{\renewcommand{\nl}{\let\nl\oldnl}}
\usepackage{csquotes}
\usepackage{hhline}
\usepackage{amssymb}
\usepackage{listings}
\usepackage{color}
\usepackage{tabularx}
\usepackage{colortbl}
\usepackage{booktabs}
\usepackage{dcolumn}
\definecolor{codegreen}{rgb}{0,0.6,0}
\definecolor{codegray}{rgb}{0.5,0.5,0.5}
\definecolor{codeamethyst}{rgb}{0.6, 0.4, 0.8}
\definecolor{ao}{rgb}{0.0, 0.0, 1.0}
\definecolor{azure(colorwheel)}{rgb}{0.0, 0.5, 1.0}
\definecolor{backcolour}{rgb}{0.95,0.95,0.92}
\usepackage[table]{xcolor}

\usepackage{subcaption}
\usepackage{titlesec}

\usepackage{dcolumn}
\usepackage{tabularx}
\usepackage{hyperref}
\hypersetup{
    colorlinks=true,
    linktoc=all,
    linkcolor=black,
    citecolor=blue,
    allcolors=ao
}
\usepackage{longtable}
\usepackage{braket}
\newcolumntype{C}{>{\centering\arraybackslash}X}
\SetAlFnt{\small}
\usepackage[font=small]{caption}
\begin{document}

\title{Solving The Vehicle Routing Problem  via Quantum Support Vector Machines}

\author{\IEEEauthorblockN{1\textsuperscript{st}Nishikanta Mohanty,}
\IEEEauthorblockA{\textit{Centre for Quantum Software and Information, } \\
\textit{University of Technology Sydney,} Ultimo, Sydney 2007, NSW, Australia \\ Nishikanta.M.Mohanty@student.uts.edu.au}\\
\and
\IEEEauthorblockN{2\textsuperscript{nd} Bikash~K.~Behera,}
\IEEEauthorblockA{\textit{Bikash's Quantum (OPC) Pvt. Ltd.,} Mohanpur 741246, WB, India\\
bikas.riki@gmail.com}\\
\and
\IEEEauthorblockN{3\textsuperscript{rd} Christopher Ferrie,}
\IEEEauthorblockA{\textit{Centre for Quantum Software and Information,} \\
\textit{University of Technology Sydney,} Ultimo, Sydney 2007, NSW, Australia \\
Christopher.Ferrie@uts.edu.au}}

%\author{Nishikanta Mohanty\thanks{N. Mohanty is Centre for Quantum Software and Information, University of Technology Sydney, Ultimo, NSW 2007 Sydney, Australia, e-mail: (Nishikanta.M.Mohanty@student.uts.edu.au)}, Bikash~K.~Behera\thanks{B.~K. Behera is associated with the Bikash's Quantum (OPC) Pvt. Ltd., Mohanpur, WB, 741246 India, e-mail: (bikas.riki@gmail.com).} and Christopher Ferrie\thanks{C. Ferrie is associated with the Department of Physics and Computer Science, Wilfrid Laurier University, Waterloo, Canada e-mail: (Christopher.Ferrie@uts.edu.au).}}

\maketitle

\begin{abstract}
The Vehicle Routing Problem (VRP) is an example of a combinatorial optimization problem that has attracted academic attention due to its potential use in various contexts. VRP aims to arrange vehicle deliveries to several sites in the most efficient and economical manner possible. Quantum machine learning offers a new way to obtain solutions by harnessing the natural speedups of quantum effects, although many solutions and methodologies are modified using classical tools to provide excellent approximations of the VRP. In this paper, we implement and test hybrid quantum machine learning methods for solving VRP of $3$ and $4$-city scenarios, which use $6$ and $12$ qubit circuits, respectively. The method is based on quantum support vector machines (QSVMs) with a variational quantum eigensolver on a fixed or variable ansatz. Different encoding strategies are used in the experiment to transform the VRP formulation into a QSVM and solve it. Multiple optimizers from the IBM Qiskit framework are also evaluated and compared.
\end{abstract}

\begin{IEEEkeywords}
Vehicle Routing Problem, Ising Model,  Variational Quantum Eigensolver, Quantum Encoding, Quantum Support Vector Machine, Parameterized Circuit
\end{IEEEkeywords}

\maketitle

\section{Introduction}

\subsection{Quantum Computing}
Quantum computing has provided novel approaches for solving computationally complex problems over the last decade by leveraging the inherent speedup(s) of quantum calculations compared to classical computing. Quantum superposition and entanglement are two key factors that give a massive speed up to calculations in the quantum domain compared to classical counterparts \cite{montanaro_quantum_2016,jordan_httpsquantumalgorithmzooorg_nodate, RevModPhys.81.865}. Because of this, addressing Optimization problems by quantum computing is an appealing prospect. Multiple approaches, such as Grover's algorithm \cite{grover_fast_1996}, adiabatic computation (AC) \cite{farhi_quantum_2000}, and quantum approximate optimization algorithm (QAOA)  \cite{farhi_quantum_2014}, have been proposed to use quantum effects and, as such, have served as the basis for solving mathematically complex problems using quantum computing. The performance of classical algorithms has generally been found to be subpar when applied to larger dimensional problem spaces \cite{nac_chapter3_quantum_algo}. On a multidimensional problem, classical machine learning optimization techniques frequently require a significant amount of CPU and GPU resources and take a long time to compute. The reason for this is because ML techniques are needed to resolve NP-hard optimization problems \cite{dasari_solving_2020}.  

\subsection{Vehicle Routing Problem}
The vehicle routing problem is an intriguing optimization problem because of its many uses in routing and fleet management \cite{ harwood_formulating_2021}, but its computational complexity is NP-hard  \cite{Kumar2012,Mohanty21}. Moving automobiles as quickly and cheaply as feasible is always the objective. VRP has inspired a plethora of precise and heuristic approaches \cite{harwood_formulating_2021,srinivasan_efficient_2018}, all of which struggle to provide fast and trustworthy solutions. The VRP's bare bones implementation comprises sending a single vehicle to deliver items to many client locations before returning to the depot to restock \cite{feld_hybrid_2019}. By optimizing a collection of routes that are available and all begin and terminate at a single node called the depot, VRP seeks to maximize the reward, which is often the inverse of the total distance traveled or the average service time. It is computationally difficult to find an optimum solution to this issue, even with just a few hundred customer nodes.

Explicitly, in every VRP ($n, k$), there are ($n-1$) stations, $k$ vehicles, and a depot D \cite{utkarsh_solving_2020, harwood_formulating_2021}. The solution is a collection of paths whereby each vehicle takes exactly one journey, and all $k$ vehicles start and conclude at the same location, $D$.
 The best route is one that requires $k$ vehicles to drive the fewest total miles. This problem may be thought of as a generalization of the well-known ``traveling salesman" problem, whereby a group of $k$ salesmen must service an aggregate of ($n-1$) sites with a single visit to each of those places being guaranteed \cite{harwood_formulating_2021}. In most practical settings, the VRP issue is complicated by other constraints, such as limited vehicle capacity or limited time for coverage. As a consequence, several other approaches, both classical and quantum, have been proposed as potential ways forwards. Currently, available quantum approaches for optimizing a system include the Quantum Approximate Optimization Algorithm (QAOA) \cite{utkarsh_solving_2020}, the Quadratic Unconstrained Binary Optimization (QUBO) \cite{glover_quantum_2020}, and quantum annealing \cite{irie_quantum_2019, crispin_quantum_2013, fujitsu_annealer_2019}.

\subsection{Quantum Support Vector Machine(QSVM)} 
The goal of the support vector machine (SVM) technique is to find the best line (or decision boundary) between two classes in $n$-dimensional space so that new data may be classified quickly. This optimum decision boundary is referred to as a hyperplane. The most extreme vectors and points that help construct the hyperplane are selected using SVM. The SVM method is based on support vectors, which are used to represent these extreme instances. Typically, a hyperplane cannot divide a data point in its original space. In order to find this hyperplane, a nonlinear transformation is applied to the data as a function. A feature map is a function that transforms the features of provided data into the inner product of data points, also known as the kernel \cite{Havlicek2019, Rebentrost_2014, kariya2021_SVM}.

Quantum computing produces implicit calculations in high-dimensional Hilbert spaces using kernel techniques by physically manipulating quantum systems. Feature vectors for SVM in the quantum realm are represented by density operators, which are themselves encodings of quantum states. The kernel of a quantum support vector machine (QSVM) is made up of the fidelities between different feature vectors, as opposed to a classical SVM; the kernel conducts an encoding of classical input into quantum states  \cite{Havlicek2019,Leporini2021}.

\subsection{Novelty and Contribution}

\begin{itemize}
    \item In this work, we propose a new method to solve the VRP using a machine-learning approach through the use of QSVM. 
    \item In this context, we came across recent and older works in QSVM \cite{kariya2021_SVM,gentinetta2022_QSVM,Rebentrost_2014} and VQE algorithms \cite{cerezo_variational_2021}, which are used to solve optimization problems such as VRP. However, none of them use a hybrid approach to arrive at a solution. 
    \item Our work implements this new approach of solving VRP in a detailed gate-based simulation of a $3$-city or $4$-city problem on a 6-qubit or 12-qubit system, respectively, using a  parameterized circuit that is developed as a solution to VRP. 
    \item We apply quantum encoding techniques such as amplitude encoding, angle encoding, higher order encoding, IQP Encoding, and quantum algorithms such as  QSVM, VQE, and QAOA to construct circuits for VRP and analyze the effects and consolidate our findings.
    \item We evaluate our solution using a variety of classical optimizers, as well as fixed and variable Hamiltonians to draw statistical conclusions.
\end{itemize}

\subsection{Organization}
The paper is organized as follows. Sec. \ref{Background} discusses the fundamental mathematical concepts such as  QAOA,  the Ising model, quantum support vector machine, Amplitude encoding, Angle encoding, Higher order encoding, IQP encoding, and VQE. Sec. \ref{Methodology} discusses the formulation and solution of VRP using the concepts discussed in the previous Section. Sub-Sec. \ref{Analysis And Circuit Building} covers the basic building blocks of circuits to solve VRP using QSVM. Sec. \ref{Results} covers the outcomes of the QSVM simulation consisting of two sub-sections. Sub Sec \ref{QSVM Simulation results} covers the outcome of simulation results of all the encoding schemes used, Finally in Sub Sec. \ref{Inferences}, we conclude by comparing the results of QSVM solutions using various optimizers in the Qiskit platform on the VRP circuit and discuss the feasibility of higher qubit solutions as the future directions of research.

\section{Background} \label{Background}
Dealing with methods and processes for resolving combinatorial optimization problems is the foundation of solving routing challenges. The objective function is then created by transforming the mathematical models into a quantum equivalent mathematical model. By maximizing or minimizing the mathematical model iteratively, we arrive at the solution of the objective function. We list the main ideas in this section for our solution approach. 

\subsection{QAOA} \label{QAOA}

A variational approach called the Quantum Approximate Optimization Algorithm (QAOA) was put forth by Farhi \emph{et al.} in 2014 \cite{farhi_quantum_2000, farhi_quantum_2014} using adiabatic quantum computation framework as the foundation of this algorithm. It is a hybrid algorithm since it applies both classical and quantum approaches. Simply described, quantum adiabatic computation involves switching from the eigenstate of the driver Hamiltonian to that of the problem Hamiltonian. The problem Hamiltonian can be expressed as,
\begin{eqnarray}
C|z\rangle =\sum^{m}_{\alpha=1}C_{\alpha}|z\rangle.
\end{eqnarray}
We are aware that the combinatorial optimization problem is resolved by finding the highest energy eigenstate of C. Similarly, we employ driver Hamiltonian as 
\begin{eqnarray}
B=\sum^n_{j=1}{\ }{\sigma}^x_j ,
\end{eqnarray}
where ${\sigma}^x_j$ represents the ${\sigma}^x$ Pauli operator on bit $z_j$ and $B$ is the mixing operator. Let's additionally define $U_C\left(\gamma\right){=}e^{{-}i\gamma C}$ and $U_B\left(\beta\right){=}e^{{-}i\beta{B}}$ which allow the system to evolve under C for $\gamma$ time and under B for $\beta $ time, respectively. Essentially, QAOA creates a state
\begin{equation}
|\boldsymbol{\beta},\boldsymbol{\gamma}\rangle =e^{-i{\beta}_pB}e^{-i{\gamma}_p{C}}\cdots e^{-i{\beta}_2B}e^{-i{\gamma}_2{C}}e^{-i{\beta}_1{B}}e^{-i{\gamma}_1{C}}|s\rangle,
\end{equation}
where $|s\rangle $ denotes the superposition state of all input qubits The expectation value of the cost function $\sum^m_{\alpha{=1}}{{\langle }\beta{,}\gamma~{|C_\alpha}~{|}\beta{,}\gamma~{\rangle }}$ gives the solution, or an approximate solution to the problem \cite{zhou_quantum_2020}.

\subsection{Ising Model}
 
 In statistical mechanics, the Ising model is a well-known mathematical depiction of ferromagnetism\cite{singh_Ising_2020,RevModPhys.39.883}. In the model, discrete variables ($+1$ or $-1$) represent the magnetic dipole moments of "spins" in one of two states. Because the spins are organized in a network, commonly a lattice(when there is periodic repetition in all directions of the local structure), each spin can interact with its neighbors.  The spins interact in pairs, with an energy that has one value when the two spins are identical and a second value when they are dissimilar. Nevertheless, heat reverses this tendency, enabling alternate structural phases to arise. The model is a condensed representation of reality that enables the recognition of phase transitions. The following Hamiltonian explains the total spin energy:

\begin{eqnarray}
H_c=-\sum_{\left\langle i,j\right\rangle }{\ }J_{ij}{{\sigma}}_{{i}}{{\sigma}}_{{j}}-h\sum {{\sigma}}_{{i}},
\end{eqnarray}
where $J_{ij}$ represents the interaction of adjacent spins $i$ and $j$, and $h$ represents an external magnetic field. The ground state at $h=0$ is a ferromagnet if $J$ is positive. If $J$ is negative, the ground-state at $h=0$ is an anti-ferromagnet for a bipartite lattice. As a result, for the sake of simplicity and in the context of this document, we can write the Hamiltonian as

\begin{eqnarray}
H_c=-\sum_{\langle i,j\rangle }{\ }J_{ij}{\sigma}_{i}^z{\sigma}_{j}^z-\sum h_i{\sigma}_{i}^x.
\end{eqnarray}

Here ${\sigma}_z$ and ${\sigma}_x$ represent Pauli $z$ and $x$ operator. For the sake of simplification, we can assume the following conditions to be ferromagnetic ($J_{ij}>0$) if there is no external impact on the spin: $h=0$. Hence, the Hamiltonian may be rewritten as follows:

\begin{eqnarray}
H_c=-\sum_{\langle i,j\rangle }{\ }J_{ij}{\sigma}_{i}^z{\sigma}_{j}^z=-\sum_{\langle i,j\rangle }{\ }\sigma_{i}^z\sigma_{j}^z.
\end{eqnarray}

\subsection{Quantum Support Vector Machine} 
SVM \cite{Rebentrost_2014,kariya2021_SVM} is a supervised algorithm that constructs hyper-plane with $\Vec{w} \cdot \Vec{x} + b = 0$ such that $\Vec{w}\cdot\Vec{x} + b \geq 1$ for a training point $\Vec{x}_i$ in the positive class, and $\Vec{w}\cdot \Vec{x} + b \leq -1 $ for a training point $\Vec{x}_i$ in the negative class. During the training process, the algorithm aims to maximize the gap between the two classes, which is intuitive as we want to separate two classes as far as possible, in order to get a sharper estimate for the classification result of new data samples like $\Vec{x_0}$. Mathematically we can see the objective of SVM is to find a hyper-plane that maximizes the distance $2/ |\Vec{w}|$ constraint to  $\Vec{y_i}(\Vec{w}\cdot \Vec{x_i} + b) \geq 1$. The normal vector $\Vec{w}$ can be written as $\Vec{w} = \sum_{i=1}^{M} \alpha_i \Vec{x}_i$ where $\alpha_i$ is the weight of the $i^{th}$ training vector $\Vec{x}_i$. Thus, obtaining optimal parameters b and $\alpha_i$ is the same as finding the optimal hyper-plane. To classify the new vector, is analogous to knowing which side of the hyper-plane it lies, i.e., $ y_i(\Vec{x}_0) = sign(\Vec{w}.\Vec{x} + b) $. After having the optimal parameters, classification now becomes a linear operation. From the least-squares approximation of SVM, the optimal parameters can be obtained by solving a linear equation,

\begin{eqnarray}
\Vec{F}(b,\alpha_1,\alpha_2,\alpha_3,...,\alpha_M)^T = (0,y_1,y_2,y_3,...y_M)^T.
\label{QSVM_Eq1}
\end{eqnarray}

In a general form of F we adopt the linear kernels $K_{i,j} = \kappa(\Vec{x}_i,\Vec{x}_j) = \Vec{x}_i. \Vec{x}_j $. Thus to find the hyper-plane parameters we use matrix inversion of F : $(b,{\vec{\alpha}_i}^T)^T  = {\tilde F^{ - 1}} (0,{\vec{y}_i}^T)^T$. 

\subsubsection{Quantum Kernels}
The main inspiration of a quantum Support vector machine is to consider quantum feature maps that lead to quantum kernel functions, which are hard to simulate in classical computers. In this case, the quantum computer is only used to estimate a quantum kernel function, which can be later used in kernel-based algorithms. For simplicity assuming the datapoints $x,z \in \mathcal{X}$, the nonlinear feature map of any datapoint is

\begin{eqnarray}
\Phi(\boldsymbol{x})=U(\boldsymbol{x})\left|0^n\right\rangle\left\langle 0^n\right| U^{\dagger}(\boldsymbol{x}) .
\end{eqnarray}

The kernel function $\kappa(x, z)$ can be computed as 
\begin{eqnarray}
\kappa(x, y)=|\langle\phi(x) \mid \phi(z)\rangle|^2.
\end{eqnarray}
The state $|\phi(x)\rangle$ can be prepared by using a unitary gate $U(x)$, and thus $|\phi(x)\rangle=U(x)|0\rangle$.Thus the kernel fuction becomes ,
\begin{eqnarray}
\kappa(x, z)=\left|\left\langle 0\left|U^{\dagger}(x) U(z)\right| 0\right\rangle\right|^2.
\end{eqnarray}

\begin{figure}[H]
    \centering
    \includegraphics[scale=0.25]{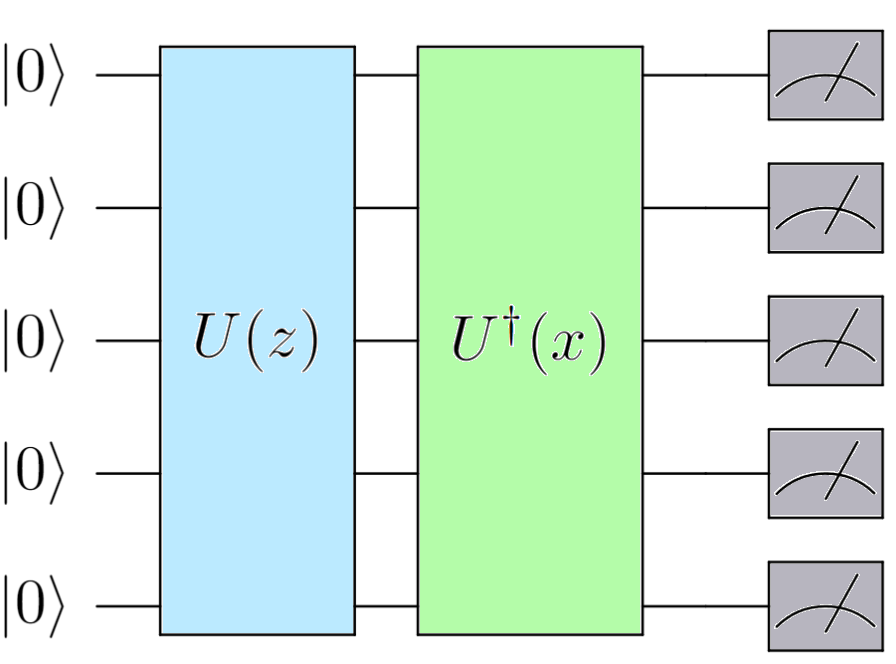}
    \caption{Schematic diagram depicting quantum circuit for Kernel estimation.}
    \label{Kernel_estimation}
\end{figure}

From the above we can say that the kernel $\kappa(x, z)$ is simply the probability of getting an all-zero string when the circuit $U^{\dagger}(x) U(z)|0\rangle$ is measured or this kernel is an  $\left|0^n\right\rangle$ to $\left|0^n\right\rangle$ transition probability of a particular unitary quantum circuit on $n$ qubits \cite{Havlicek2019,glick_covariant_kernel_2022}. This can be implemented using the following kernel estimation circuit (Fig. \ref{Kernel_estimation}).

\subsection{Amplitude Encoding(AE)}

In the process of amplitude-embedding \cite{araujo_divide-and-conquer_2021}, data is encoded into the amplitudes of a quantum state. A N-dimensional classical datapoint $x$ is represented by the amplitudes of an n-qubit quantum state $\ket {\psi_x}$ as  

\begin{equation}
\left|\psi_x\right\rangle=\sum_{i=1}^N x_i|i\rangle
\end{equation}
where $N=2^n$, 
$x_i$   is the $i$-th element of $x$  
and $|i\rangle$ is the  $i$-th computational basis state.

In order to encode any data point $x$ into an amplitude-encoded state, we must normalize the same  by following

\begin{equation}
\ket{\psi_{x_{norm}}}=\frac{1}{x_{norm}}\sum_{i=1}^N x_i|i\rangle ,
\end{equation}

where $x_{norm}$= $\sqrt{\sum_{i=1 }^N|x_i|^2}$ .

\subsection{Angle Encoding(AgE)}

While the above-described amplitude encoding expands into a complicated quantum circuit with huge depths, the angle encoding employs N qubits and a quantum circuit with fixed depth, making it favorable to NISQ computers \cite{LaRose_dataencodings_2020,qiskit_lecture_2021}. We define angle encoding as a method of classical information encoding that employs rotation gates(the rotation could be chosen along $x$, $y$ or $z$ axis). In our scenario, the classical information consists of the node and edge weights assigned to the vehicle's nodes and pathways which are further assigned as parameters to ansatz.

\begin{equation}
|\mathbf{x}\rangle=\bigotimes_i^n R\left(\mathbf{x}_i\right)\left|0^n\right\rangle,
\end{equation}

where ${x}_i$ represents the classical information stored on the angle parameter of rotation operator $R$.

\subsection{ Higher Order Encoding(HO)}

Higher order encoding is a variation of angle encoding where we have an entangled layer and an additional sequential operation of rotation angles of two entangled qubits \cite{qiskit_lecture_2021}. This can be loosely defined as the following

\begin{equation}
|\mathbf{x}\rangle=\bigotimes_{i=2}^{n}R_(x_{i-1}.x_i)\bigotimes_{i=2}^{n-1}CX_{i,i+1}\bigotimes_{i=1}^n R\left(x_i\right)\left|0^n\right\rangle.
\end{equation}

Similar to angle encoding we are free to chose the rotation.

\subsection{IQP Encoding(IqpE)}
IQP-style encoding is a relatively complicated encoding strategy. We encode classical information \cite{noauthor_-paddle_nodate}

\begin{equation}
|{x}\rangle=\left(\mathrm{U}_{\mathrm{Z}}({x}) \mathrm{H}^{\otimes n}\right)^r\left|0^n\right\rangle,
\end{equation}

where $r$ is the depth of the circuit, indicating the repeating times of $\mathrm{U}_{\mathrm{Z}}({x}) \mathrm{H}^{\otimes n}$. $\mathrm{H}^{\otimes n}$ is a layer of Hadamard gates acting on all qubits. $\mathrm{U}_{\mathrm{Z}}(\mathbf{x})$ is the key step in IQP encoding scheme:

\begin{eqnarray}
\mathrm{U}_{\mathrm{Z}}(\mathrm{x})=\prod_{[i, j] \in S} R_{Z_t Z_j}\left(x_i x_j\right) \bigotimes_{k=1}^n R_z\left(x_k\right),
\end{eqnarray}
where $S$ is the set containing all pairs of qubits to be entangled using $R_{Z Z}$ gates.
First, we consider a simple two-qubit gate: $R_{Z_1 Z_2}(\theta)$. Its mathematical form $e^{-i \frac{i}{2} Z_1 \otimes Z_2}$ can be seen as a two-qubit rotation gate around $Z Z$, which makes these two qubits entangled.

\subsection{VQE}

Another hybrid quantum classical algorithm is the Variational Quantum Eigensolver (VQE), which is used to estimate the eigenvalue of a large matrix or Hamiltonian $H$ \cite{peruzzo_variational_2014}. The basic goal of this method is to find a trial qubit state of a wave function $\ket{\psi(\vec\theta)}$ that is dependent on a parameter set $\vec\theta = \theta_1,\theta_2,\cdots$, which is also known as the variational parameters. The expectation of an observable or Hamiltonian $H$ in a state  $\ket{\psi(\vec\theta)}$ can be expressed in quantum theory as,

\begin{eqnarray}
E(\vec{\theta})= \bra{\psi(\vec\theta)} H \ket{\psi(\vec\theta)}.
\label{Eq. Expectation}
\end{eqnarray} 

By spectral decomposition 
\begin{eqnarray}
H=\lambda_{1}\ket\psi_{1}\bra\psi_{1}+\lambda_{2}\ket \psi_{2}\bra\psi_{2}+\ldots+\lambda_{n}\ket \psi_{n}\bra\psi_{n},\nonumber\\
\end{eqnarray}

where $\lambda_i$ and ${\ket\psi}_i$ are the eigenvalues and eigenstates,
respectively, of matrix $H$. Also, because the eigenstates of $H$ are orthogonal, $\left\langle\psi_{i} \mid \psi_{j}\right\rangle=0$  If $i \neq j$ . The wave function $\ket{\psi(\vec\theta)}$ can be expressed as a superposition of eigenstates.

\begin{eqnarray}
\ket{\psi(\vec\theta)}=\alpha_{1}(\vec\theta)\ket\psi_{1}+\alpha_{2}(\vec\theta)\ket\psi_{2}+\ldots+\alpha_{n}(\vec\theta)\ket\psi_{n}.
\end{eqnarray}

Thus the expectation becomes 
\begin{eqnarray}
E(\vec\theta )  &=& |\alpha_{1}(\vec\theta ) |^2 \lambda_{1}+|\alpha_{2} (\vec\theta ) |^2\lambda_2+\ldots+|\alpha_{n}(\vec\theta ) |^2 \lambda_{n}.\nonumber\\
\end{eqnarray}

Clearly, $ E(\vec\theta) \geq \lambda_{\min }$.
So in VQE algorithm, we vary the parameters $\vec{\theta}=\theta_{1}, \theta_{2}, \ldots$ until $E(\vec{\theta})$ is minimized. This property of VQE is useful when attempting to solve combinatorial optimization problems namely those in which a parameterized circuit is used to set up the trial state of the algorithm, and $E(\vec{\theta})$ is referred to as the cost function, that is also the expected value of the Hamiltonian in this state. The ground state of the desired Hamiltonian may be obtained by iterative minimization of the cost function.The optimization process utilizes a classical optimizer which uses quantum computer to evaluate the cost function and calculate its gradient at each optimization step.

\section{Methodology} \label{Methodology}

\subsection{Modelling VRP in QSVM} \label{Modelling QSVM}

The vehicle routing problem can be solved by mapping the cost function to an Ising Hamiltonian $H_c$ \cite{lucas_Ising_2014}. The answer to the problem is given by minimizing the Ising Hamiltonian $H_c$. Consider an arbitrarily connected graph with $n$ vertices and $n-1$ edges. Assuming we need to route a vehicle between two non-adjacent vertices in the graph; Consider a binary decision variable $x_{ij}$ whose value is $1$ if there is an edge between $i$ and $j$ with an edge weight $w_{ij}>0$; otherwise, its value is $0$. Now, the VRP problem requires $n \times (n-1)$ choice variables. We define two sets of nodes for each edge from $i\rightarrow j$: $source\left[i\right]$ and $target[j]$. $source\left[i\right]$ contains the nodes $j$ to which $i$ sends an edge $j\ \epsilon\ source [i]$. The collection $target\left[j\right]$ comprises the nodes $i$ to which the node $i$ delivers the edge $i\ \epsilon\ target[j]$. The VRP is defined as follows\cite{utkarsh_solving_2020,Qiskit_Optimization_VRP}:

\begin{eqnarray} 
VRP(n,k)=\mathop{min}_{{\left\{x_{ij}\right\}}_{i\to j}\in \{0,1\}}\ \sum_{i\to j}{\ }w_{ij}x_{ij},
\end{eqnarray}

where $k$ is the number of vehicles, and $n$ is the total number of locations, we have $n-1$ locations for vehicles to traverse if we consider the starting place to be the $0th$ location or Depot $D$. Noticeably, this is subject to the following restrictions\cite{Mohanty21}:

\begin{eqnarray}
\sum_{j\in ~{source}~[i]}{\ }x_{ij}&=&1,{\forall }i\in \{1,\cdots ,n-1\} , \nonumber\\
\sum_{j\in ~{target}~[i]}{\ }x_{ji}&=&1,{\forall }i\in \{1,\cdots ,n-1\} , \nonumber\\
\sum_{{j}{\in }~{source}~{[}{0}{]}}{{\ }}{x}_{0j}&=& k, \nonumber\\
\sum_{j\in ~{target}~[0]}{\ }x_{j0}&=&k \nonumber\\
u_i-u_j+Q x_{i j} &\leq& Q-q_j, \forall i \sim j, i, j \neq 0, \nonumber\\
 q_i \leq u_i &\leq& Q, \forall i, i \neq 0. 
\end{eqnarray}

The first two restrictions establish the limitation that the delivering vehicle may only visit each node once. After delivering the products, the middle two limitations enforce the requirement that the vehicle must return to the depot. 
The last two constraints impose the sub-tour elimination conditions and are bound on $u_i$, with $Q>q_j>0$, and $u_i,Q, q_i \in \mathbb{R}$. For the VRP equation and restrictions, the Hamiltonian of VRP can be expressed as follows \cite{utkarsh_solving_2020}.

\begin{eqnarray}
H_{VRP}&=&H_A+H_B+H_C+H_D+H_E, \nonumber\\
H_A&=&~\sum_{i~\to j}{w_{ij}x_{ij}} ,\nonumber\\
H_B&=&A\sum_{i\in 1,\cdots ,n-1}{\ }{\left(1-\sum_{j\in ~{source}~[i]}{\ }x_{ij}\right)}^2, \nonumber\\
H_C&=&A\sum_{i\in 1,\cdots ,n-1}{\ }{\left(1-\sum_{j\in ~{target}[i]}{\ }x_{ji}\right)}^2, \nonumber\\ 
H_D&=&A{\left(k-\sum_{j\in ~{source}[0]}{\ }x_{0j}\right)}^2, \nonumber\\
H_E&=&A{\left(k-\sum_{j\in ~{target}[0]}{\ }x_{j0}\right)}^2 .
\label{Eq. VRP}
\end{eqnarray}

$A > 0$ represents a constant. In vector form, the collection of all binary decision variables $x_{ij}$  can be written as

\begin{eqnarray}
\overrightarrow{\boldsymbol{{x}}}={\left[x_{(0,1)},x_{(0,2)},\cdots x_{(1,0)},x_{(1,2)},\cdots x_{(n-1,n-2)}\right]}^{\boldsymbol{{T}}} .
\end{eqnarray}

Using the preceding vector, we can build two new vectors for each node: $\overrightarrow{z}_{S[i]}$ and $\overrightarrow{z}_{T[i]}$ (in the beginning of the section, we defined two sets for source and target nodes, thus two vectors will represent them).

\begin{eqnarray}
\overrightarrow{z}_{S\left[i\right]}&=& \vec{x} \ni x_{ij}=1,\ x_{kj}=0\ ,\ k\neq i\ ,\ \ \forall j,k\ \in \{0,\cdots ,n-1\} , \nonumber\\ 
\overrightarrow{z}_{T\left[i\right]}&=&\vec{x} \ni x_{ji}=1,\ x_{jk}=0\ ,\ k\neq i\ ,\ \ \forall j,k\ \in \{0,\cdots ,n-1\} . \nonumber\\
\end{eqnarray}
\begin{eqnarray}
\sum_{j \in \text { source }[i]} x_{i j}&=&\vec{z}_{S[i]}^{\mathrm{T}} \vec{x} , \nonumber\\
\sum_{j \in \text { target }[i]} x_{j i}&=&\vec{z}_{T[i]}^{\mathrm{T}} \vec{x} . 
\label{NEVec}
\end{eqnarray}

The aforementioned vectors will aid in the development of the QUBO model of VRP \cite{date_efficiently_2019, glover_quantum_2020, kochenberger_unconstrained_2014, guerreschi_solving_2021}. In general, the QUBO model of a connected graph $G=(N,V)$ is specified as follows:

\begin{eqnarray}
f(x)_{QUBO}=\mathop{min}_{x\in \{0,1\}(N\times V)}x^TQx+g^Tx+c,
\label{Eq. QUBO}
\end{eqnarray}

where, $Q$ is a quadratic edge weight coefficient, $g$ is a linear node weight coefficient, and $c$ is a constant. In order to find these coefficients in the QUBO formations of $H_{VRP}$ given in Eq. \ref{Eq. VRP} we first put in Eqs. \ref{NEVec} in terms $H_{B}$  and $H_{c}$respectively, then expand and regroup the expression of $H_{VRP}$ according to Eq. \ref{Eq. QUBO}

\begin{eqnarray}
%\begin{aligned}
H&=&A \sum_{i=0}^{n-1}\left[z_{S[i]} z_{S[i]}^T+z_{T[i]} z_{T[i]}^T\right] \vec{x}^2 \nonumber \\
& +&w^T \vec{x}-2 A \sum_{i=1}^{n-1}\left[z_{S[i]}^T+z_{T[i]}^T\right] \vec{x} \nonumber\\
&-& 2 A k\left[z_{S[0]}^T+z_{T[0]}^T\right] \vec{x}+2 A(n-1)+2 A k^2.\nonumber\\
%\end{aligned}
\end{eqnarray}

Hence for QUBO formulation of Eq. \eqref{Eq. VRP} we get the coefficients $\mathrm{Q}(n(n-1) \times$ $n(n-1)), \mathrm{g}(n(n-1) \times 1)$ and $\mathrm{c}$ :  The coefficients for the QUBO formulation of Eq. \eqref{Eq. VRP} are therefore as follows:

\begin{eqnarray}
Q&=&A\left[\left[z_{T[0]}, \ldots, z_{T[n-1]}\right]^{T}\left[z_{T[0]}, \ldots, z_{T[n-1]}\right]\right. \nonumber\\
&&\left.+\left(\mathbb{I}_{n} \otimes \mathbb{J}(n-1, n-1)\right)\right], \nonumber\\
g&=&W-2 A k\left(\left(e_{0} \otimes \mathbb{J}_{n-1}\right)+\left[z_{T[0]}\right]^{T}\right), \nonumber\\
\quad&&+2 A\left(\mathbb{J}_{n} \otimes \mathbb{J}_{n-1}\right), \nonumber\\
c&=&2 A(n-1)+2 A k^{2} .
\end{eqnarray}

$\mathbb{J}$ is the matrix containing all ones, $\mathbb{I}$ is the identity matrix, and $e_0={\left[1,0,\cdots ..,0\right]}^T$ is the identity matrix.The binary decision variable $x_{ij}$ is converted to the spin variable $s_{ij}  \in \left\{-1,1\right\}$ using the formula $x_{ij}=(s_{ij}+1)/2$.

From the aforementioned equations, we may expand Eq. \eqref{Eq. QUBO} to form the Ising Hamiltonian of VRP  \cite{glover_quantum_2020}.

\begin{eqnarray}
H_{Ising}=-\sum_i{\ }\sum_{i<j}{\ }J_{ij}s_is_j-\sum_i{\ }h_is_i+d . 
\end{eqnarray}

Following are definitions for the terms $J_{ij}, h_i$, and $d$:

\begin{eqnarray}
J_{ij}&=&\ -\frac{Q_{ij}}{2},\ \forall \ i<j  ,\nonumber\\
h_i&=&\frac{g_i}{2}+\sum{\frac{Q_{ij}}{4}+\ \sum{\frac{Q_{ji}}{4}\ }\ }, \nonumber\\
d&=&c+\sum_i{\ }\frac{g_i}{2}+\sum_i{\ }\sum_j{\ }\frac{Q_{ij}}{4} .
\end{eqnarray}

\begin{figure}[!ht]
\centering
\begin{subfigure}{\linewidth}
\includegraphics[width=\linewidth]{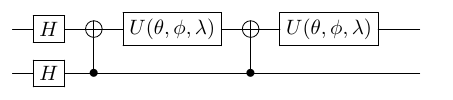} 
\caption{}
\label{qdctc_Fig3a}
\end{subfigure}\hfill
\begin{subfigure}{\linewidth}
\includegraphics[width=\linewidth]{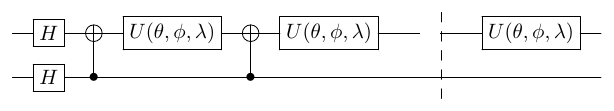} 
\caption{}
\label{qdctc_Fig3b}
\end{subfigure}\hfill
\caption{
(a) Circuit example illustrating gate operations for ${H}_{\mathrm{cost}}$. (b) Circuit example displaying gate selections with an additional $u$ gate for ${H}_{\mathrm{mixer}}$.
}
\label{qdctc_Fig3}
\end{figure}

\subsection{Analysis And Circuit Building} \label{Analysis And Circuit Building}
\subsubsection{VRP}
In this section, we create a gate-based circuit to realize the above formulation using the IBM gate model, which we have implemented using the Qiskit framework \cite{Qiskit}. For any arbitrary VRP problem using qubits, we begin with the state of $\ket{+}^{\otimes n(n-1)}$ the ground state of $H_{mixer}$ by applying the Hadamard to all qubits initialized as zero state, and we prepare the following state.

\begin{eqnarray}
\ket{\beta,\gamma}&=&e^{-iH_{mixer}\beta_{p}}e^{-iH_{cost}\gamma_{p}}...\nonumber\\
&&...e^{-iH_{mixer}\beta_{0}}e^{-iH_{cost}\gamma_{0}}\ket{+}^{n\otimes(n-1)}.
\label{eq. State}
\end{eqnarray}

The energy E of the state $\ket{\beta,\gamma}$ is calculated by expectation of $H_{cost}$ from  Eq. \eqref{Eq. Expectation}. Again From the Ising model, $H_{cost}$ term can be written in terms of Pauli operators as, 

\begin{eqnarray}
{H}_{\mathrm{cost}}=-\sum_{i} \sum_{i<j} J_{ij}\sigma_i^z\sigma_j^z-\sum_{i} h_i\sigma_i^z-d.
\end{eqnarray}

Thus for a single term of state in $\ket{\beta,\gamma}$ as $\beta_0,\gamma_0$, the expression reads,

\begin{math} 
e^{-i{H}_{mixer}\beta_0}e^{-i{H}_{cost}\gamma_0}. 
\end{math} 
The first term ${H}_{\mathrm{cost\ }}$ can be expanded to following,

\begin{eqnarray}
{e}^{iJ_{ij}\gamma_0\sigma_i\sigma_j}&=&\cos J_{ij}\gamma_0I+i\ \sin J_{ij}\gamma_0\sigma_i\sigma_j,\nonumber\\
&=& \left[\begin{matrix}{e}^{i{J}_{ij}{\gamma}_{0}} &0&0&0\\ 0&{e}^{-i{J}_{ij}{\gamma}_{0}}&0&0\\0&0&{e}^{-i{J}_{ij}{\gamma}_{0}}&0\\0&0&0&{e}^{i{J}_{ij}{\gamma}_{0}}\end{matrix}\right], \nonumber\\
&=& M
\end{eqnarray}

Applying $CNOT$ gate on before and after the above matrix `$M$' we can swap the diagonal elements,

\begin{eqnarray}
CNOT(M)CNOT=\left[\begin{matrix}{e}^{i{J}_{ij}{\gamma}_{0}} &0&0&0\\ 0&{e}^{-i{J}_{ij}{\gamma}_{0}}&0&0\\0&0&{e}^{i{J}_{ij}{\gamma}_{0}}&0\\0&0&0&{e}^{-i{J}_{ij}{\gamma}_{0}}\end{matrix}\right].\nonumber\\
\end{eqnarray}

Observing the upper and lower blocks of matrix we can rewrite,

\begin{eqnarray}
\left[\begin{matrix}1&0\\0&1\end{matrix}\right]\otimes\left[\begin{matrix}{e}^{i{J}_{ij}{\gamma}_{0}}&0\\0&{e}^{-i{J}_{ij}{\gamma}_{0}}\end{matrix}\right] = I\otimes {e}^{i{J}_{ij}{\gamma}_{0}}\left[\begin{matrix}1&0\\0&{e}^{-2i{J}_{ij}{\gamma}_{0}}\end{matrix}\right].\nonumber\\
\end{eqnarray}

$\left[\begin{matrix}1&0\\0&{e}^{-2i{J}_{ij}{\gamma}_{0}}\end{matrix}\right]$ is a phase gate. Looking at the $2$-nd term of ${H}_{\mathrm{cost}}$ we get,

\begin{eqnarray}
{H}_{\mathrm{cost}}&=&\sum_{i} h_i\sigma_{i}^{z} , \nonumber\\
e^{ih_{i}\sigma_i}&=&{\cos h}_i{\gamma_oI+i\sin\gamma_0\sigma_i},\nonumber\\
&=&\cos h_i{\gamma_o\left[\begin{matrix}1&0\\0&1\\\end{matrix}\right]+i\sin h_i\gamma_0\left[\begin{matrix}1&0\\0&-1\\\end{matrix}\right]},\nonumber\\
&=& \left[\begin{matrix}e^{ih_i\gamma_0}\ &0\\0&e^{-ih_i\gamma_0}\\\end{matrix}\right].
\end{eqnarray}

Fig. \ref{qdctc_Fig3a} depicts the basic circuit with two qubits along with gate selections for ${H}_{\mathrm{cost}}$. Similarly, $H_{mixer}$ is merely a rotation along the $X$ axis, as depicted by the $U$  gate in Fig. \ref{qdctc_Fig3b}.

The above sample circuits can be used for the solution of VRP combined with VQE and QAOA approach, However, in this paper, we are focusing on a machine learning solution of VRP by use of QSVM; thus we need to construct a QSVM circuit using various encoding schemes. Simple interpretation and implementation of encoding schemes are described in upcoming subsections.

\subsubsection{Amplitude Encoding}
As we look into AE, a single qubit state is represented by
\begin{eqnarray}
\ket\psi(\theta)=\cos(\theta/2)\ket{0}+\sin(\theta/2)\ket{1},
\end{eqnarray}

for two qubits

\begin{eqnarray}
&&|\psi(\theta)\rangle=\alpha|00\rangle+\beta|01\rangle+\gamma|10\rangle+\delta|11\rangle , \nonumber\\
&&=|0\rangle(\alpha|0\rangle+\beta|1\rangle)+|1\rangle(\gamma|0\rangle+\delta|1\rangle) , \nonumber\\
&&=|0\rangle \sqrt{\left(\alpha^2+\beta^2\right)}\left(\frac{\alpha|0\rangle+\beta|1\rangle}{\sqrt{\alpha^2+\beta^2}}\right) \nonumber\\ &&+|1\rangle \sqrt{\gamma^2+\delta^2} \frac{\gamma|0\rangle+\delta|1\rangle}{\sqrt{\gamma^2+\delta^2}} .
\end{eqnarray}

Now applying Ctrl U and Anti-CTRL U on the above state we achieve 

\begin{eqnarray}
&&|0\rangle \sqrt{\alpha^2+\beta^2}|0\rangle+|1\rangle \sqrt{\gamma^2+\delta^2}|0\rangle \nonumber\\
&&=\left(\sqrt{\alpha^2+\beta^2}|0\rangle+\sqrt{\gamma^2+\delta^2}|1\rangle\right)|0\rangle .
\end{eqnarray}

Here 
$\theta_1 = \tan^{-1} \frac{\sqrt{\gamma^2+ \delta^2}}{\sqrt{\alpha^2 + \beta^2}}$ ,
$\theta_2= \tan^{-1}\frac{\delta}{\gamma}$ ,
$\theta_3= \tan^{-1}\frac{\beta}{\alpha}$
Combining VRP and amplitude encoding circuit eliminates the need of Hadamard gates and $H_{mixer}$ components and we end up with the following skeleton circuits Fig. \ref{qdctc_Fig3} (a).

\subsubsection{Angle Encoding}

For a 2-qubit scenario, angle encoding translates to the following example. We define the $R_y$ gate as follows 
\begin{eqnarray}
R_y(\theta)&=&e^{-i Y \theta / 2}=\cos \frac{\theta}{2}-i \sin{\theta / 2}  Y, \nonumber\\
& =&\left[\begin{array}{ll}
\cos \theta / 2 & -\sin \theta / 2 \\
\sin \theta / 2 & \cos \theta / 2
\end{array}\right] .
\end{eqnarray}

\begin{eqnarray}
&& |00\rangle  \nonumber \\
&& \xrightarrow[R_y(\theta_2)]{R_y(\theta_1)}\left(\cos \frac{\theta_1}{2}|0\rangle+\sin \frac{\theta_1}{2}|1\rangle\right)\left(\cos \frac{\theta_2}{2}|0\rangle+\sin \frac{\theta_2}{2}|1\rangle\right), \nonumber\\
&& =\cos \frac{\theta_1}{2} \cdot \cos \frac{\theta_2}{2}|00\rangle+\cos \frac{\theta_1}{2} \cdot \sin \frac{\theta_2}{2}|01\rangle\nonumber \\ 
&+&\sin \frac{\theta_1}{2} \cdot \cos \frac{\theta_2}{2}|10\rangle+\sin \frac{\theta_1}{2} \sin \frac{\theta_2}{2}|11\rangle \nonumber\\
&& \xrightarrow[]{CNOT} \cos \frac{\theta_1}{2} \cdot \cos \frac{\theta_2}{2}|00\rangle+\cos \frac{\theta_1}{2} \cdot \sin \frac{\theta_2}{2}|01\rangle \nonumber\\ 
&&+\sin \frac{\theta_1}{2} \cdot \cos \frac{\theta_2}{2}|11\rangle+\sin \frac{\theta_1}{2} \sin \frac{\theta_2}{2}|10\rangle .
\end{eqnarray}

\begin{figure*}[!ht]
\centering
\begin{subfigure}{0.6\linewidth}
\includegraphics[width=\linewidth]{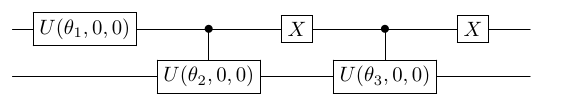} 
\caption{}
\label{qdctc_Fig3b}
\end{subfigure}\hfill
\begin{subfigure}{0.3\linewidth}
\includegraphics[width=\linewidth]{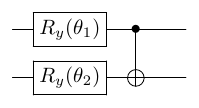} 
\caption{}
\label{qdctc_Fig3a}
\end{subfigure}\hfill
\begin{subfigure}{0.4\linewidth}
\includegraphics[width=\linewidth]{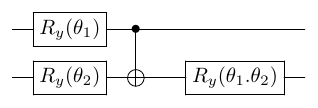} 
\caption{}
\label{qdctc_Fig3b}
\end{subfigure}\hfill
\begin{subfigure}{0.3\linewidth}
\includegraphics[width=\linewidth]{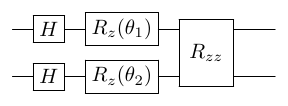} 
\caption{}
\label{qdctc_Fig3b}
\end{subfigure}\hfill
\caption{Plot illustrating different encoding methods for two qubits. (a) Amplitude encoding, (b) angle encoding, (c) Higher order encoding, (d) IQP encoding.}
\label{qdctc_Fig3}
\end{figure*}

\subsubsection{Higher Order Encoding}
For a $2$qubit scenario, HO encoding translates to the following
We define the $R_y$ gate as follows 
\begin{eqnarray}
R_y(\theta)&=&e^{-i Y \theta / 2}=\cos \frac{\theta}{2}-i \sin{\theta / 2}  Y, \nonumber\\
& =&\left[\begin{array}{ll}
\cos \theta / 2 & -\sin \theta / 2 \\
\sin \theta / 2 & \cos \theta / 2
\end{array}\right].
\end{eqnarray}

\begin{equation}
\begin{aligned}
& |00\rangle \\
& \xrightarrow[R_y(\theta_2)]{R_y(\theta_1)}\left(\cos \frac{\theta_1}{2}|0\rangle+\sin \frac{\theta_1}{2}|1\rangle\right)\left(\cos \frac{\theta_2}{2}|0\rangle+\sin \frac{\theta_2}{2}|1\rangle\right), \\
& =\cos \frac{\theta_1}{2} \cdot \cos \frac{\theta_2}{2}|00\rangle+\cos \frac{\theta_1}{2} \cdot \sin \frac{\theta_2}{2}|01\rangle \\ 
&+\sin \frac{\theta_1}{2} \cdot \cos \frac{\theta_2}{2}|10\rangle+\sin \frac{\theta_1}{2} \sin \frac{\theta_2}{2}|11\rangle \\
& \xrightarrow[R_y(\theta_1.\theta_2)]{CNOT} \cos \frac{\theta_1}{2} \cdot \cos \frac{\theta_2}{2}|0\rangle\left(\cos \frac{ \theta_1 \cdot \theta_2}{2}|0\rangle+\sin \frac{\theta_1 \cdot \theta_2}{2}|1\rangle\right)\\
&+\cos \frac{\theta_1}{2} \cdot \sin \frac{\theta_2}{2}|0\rangle\left(-\sin \frac{ \theta_1 \cdot \theta_2}{2}|0\rangle+\cos \frac{\theta_1 \cdot \theta_2}{2}|1\rangle\right) \\
& +\sin \frac{\theta_1}{2} \cos \frac{\theta_2}{2}|1\rangle\left(\cos \frac{\theta_1 \cdot \theta_2}{2}|0\rangle+\sin \frac{\theta_1 \cdot \theta_2}{2}|1\rangle\right) \\
& +\sin \frac{\theta_1}{2} \cdot \sin \frac{\theta_2}{2}|1\rangle\left(-\sin \frac{\theta_1 \cdot \theta_2}{2} |0\rangle+\cos \frac{\theta_1 \cdot \theta_2}{2}|1\rangle\right). \\
&
\end{aligned}
\end{equation}

\subsubsection{IQP Encoding}

 For a $2$qubit scenario IqpE translates to the following
\begin{equation}
\begin{aligned}
& |00\rangle \xrightarrow{H_1 H_2} |++\rangle, \\
& =\frac{1}{2}(|00\rangle+|01\rangle+|10\rangle+|11\rangle), \\
& \xrightarrow[R_Z(\theta_2)]{R_Z(\theta_1)} \frac{1}{2}\left(|00\rangle+e^{i \theta_2}|01\rangle+e^{i \theta_1}|10\rangle+e^{i\left(\theta_1+\theta_2\right)}|11\rangle\right) \\
& \left.\stackrel{\text { CNOT }}{\longrightarrow} \frac{1}{2}(|00\rangle+e^{i \theta_2}|01\rangle+e^{i \theta_1}\left|11\right\rangle+e^{i\left(\theta_1+\theta_2\right)}|10\rangle\right) \\
& \xrightarrow{R_Z(\theta_1.\theta_2)} \frac{1}{2}(|00\rangle+e^{i \theta_2(1+\theta_1)}|01\rangle+e^{i \theta_1(1+\theta_2)}|11\rangle \\
& +e^{i \theta_1(1+\theta_2)}|10\rangle+e^{i(\theta_1+\theta_2)}|11\rangle). \\
\end{aligned}
\end{equation}

\section{Results} \label{Results}

\subsection{VQE Simulation of QSVM and VRP} \label{QSVM Simulation results}

We build the Hamiltonian with a uniform distribution of weights between $0$ and $1$, and then run it along with the ansatz via IBM's three available VQE optimizers (COBYLA, L\_BFGS\_B, and SLSQP). We run the circuit up to two layers and gather data using all of the available optimizers. We run the experiment again with a fixed Hamiltonian and, subsequently, a set of variable Hamiltonians to see whether the QSVM and encoding approach can effectively reach the classical minimum. Our results indicate that COBYLA is the most efficient optimizer, followed by SLSQP and L BFGS B. In the sections that follow, we'll have a look at the results obtained using various QSVM encoding schemes. We define two terms—Accuracy and Error—in the context of outcomes' interpretability. An error occurs when the solution deviates from the classical minimum more often than it reaches it, whereas accuracy is defined as the number of times the solution reaches the classical minimum. Percentages based on the distribution of the outcomes are used to evaluate both terms.

\begin{eqnarray}
Acc &=& \frac{N}{T} , \nonumber \\
Err &=& \frac {T - N}{T}.
\end{eqnarray}

$T =$ Total number of Simulation runs  \\
$N =$ Total number of times solution reaches classical minimum \\

\subsubsection{Amplitude Encoding}

\begin{figure*}[!ht]
\centering
\begin{subfigure}{0.5\linewidth}
\includegraphics[height=3.8cm,width=8cm]{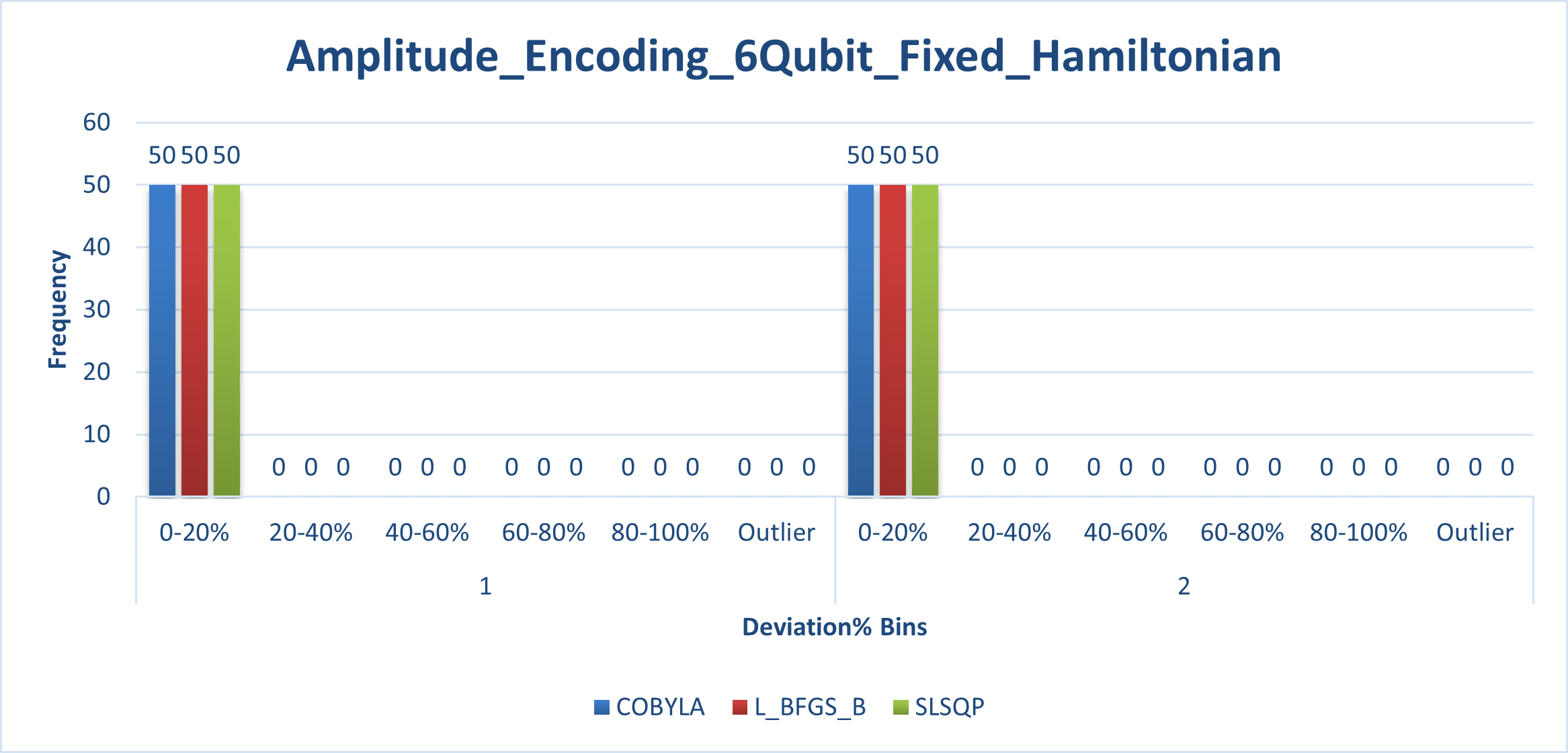} 
\caption{}
\label{}
\end{subfigure}\hfill
\begin{subfigure}{0.5\linewidth}
\includegraphics[height=3.8cm,width=8cm]{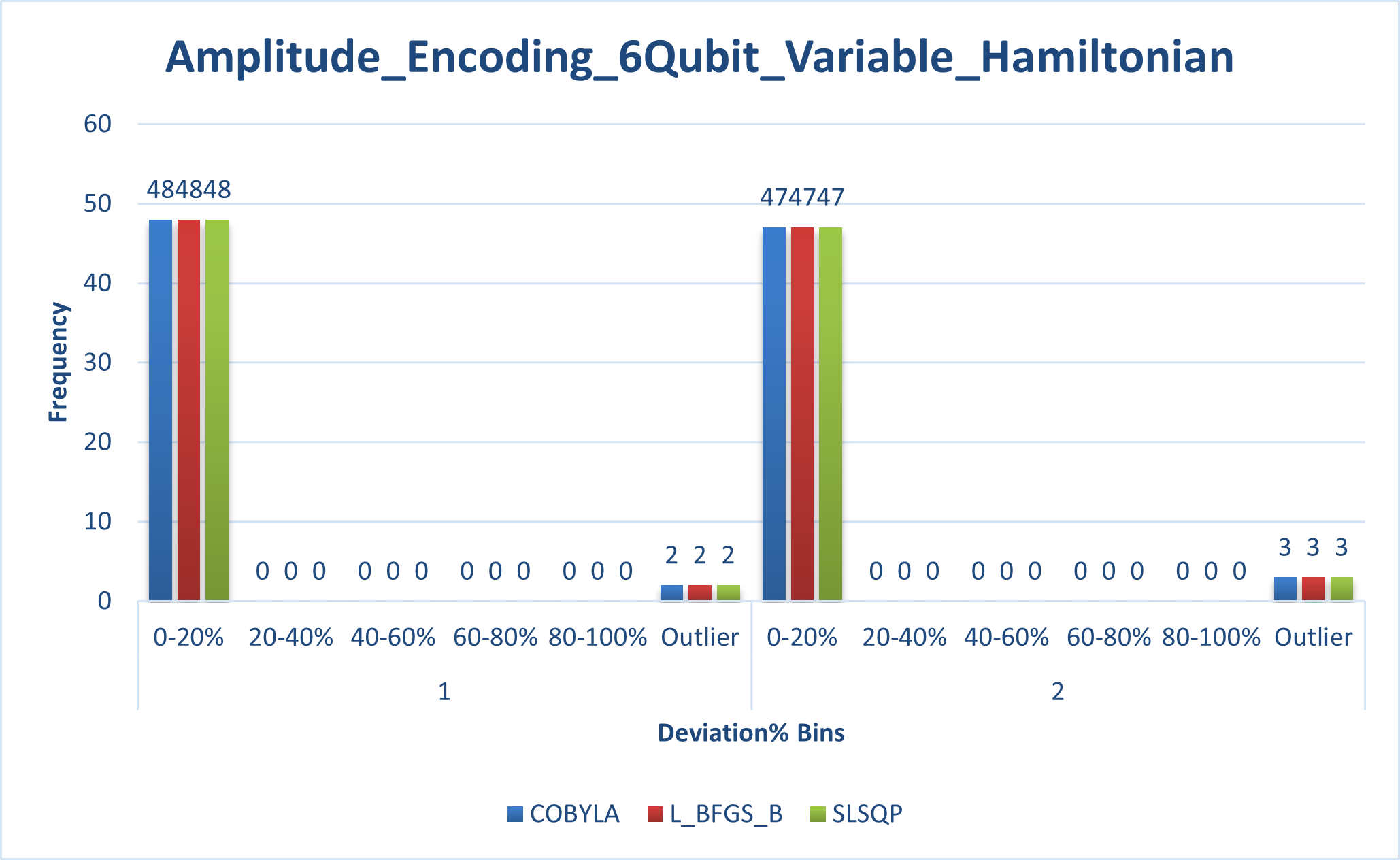} 
\caption{}
\label{}
\end{subfigure}\hfill

\caption{Plot illustrating Amplitude encoding results for QSVM solution of VRP. (a) Amplitude encoding $6$ qubits Fix Hamiltonian, (b) Amplitude encoding $6$ qubits Variable hamiltonian}
\label{AmplitudeEncodingResults}
\end{figure*}

With a large number of gates, the AE circuit has proven to be the most complex of all encoding circuits. We can simulate no more than six qubit computations due to this complexity. Despite its complexity, AE has a high, nearly perfect accuracy rate ($100\%$) and a very low error rate ($0\%$) for 50-iteration fixed Hamiltonian simulations. The trend is present in both the first and second layer. The first layer accuracy for a variable Hamiltonian simulation is $96\%$, and the second layer accuracy is $94\%$ across all optimizers. Figure \ref{AmplitudeEncodingResults} depicts the results of $50$ iterations of simulating SVM with amplitude encoding on a VRP circuit with fixed and variable Hamiltonian. The decline in accuracy, however, can be attributed to simulation or computational errors, as all the errors are greater than $100$ percent and are therefore considered aberrations. Most likely, the simulation hardware cannot accommodate the VQE procedure.

\subsubsection{Angle Encoding}

Angle encoding is the second encoding, following amplitude encoding; we have experimented with SVM VRP simulation, which yields high accuracy and low error rates. Observing tables I and II, angle encoding is the second most precise encoding employed in our investigations. For fixed Hamiltonian simulations over $50$ iterations with $6$ qubits angle encoding, the first layer, including all optimizers, achieves $100$ percent accuracy and zero percent error. In the 2nd layer simulation (over $50$ iterations), the accuracy decreases to $98\%$ for COBYLA, $96\%$ for SLSQP, and $86\%$ for L\_BGFS\_B, which is a greater decrease than the other two. These declines are attributable to optimizer-dependent statistical errors. Similarly, for $12$ qubit simulations of SVM VRP, the accuracy rates are higher in the first layer, which consists of COBYLA at $100\%$, SLSQP at $92\%$, and L\_BGFS\_B at $88\%$, reiterating that the accuracy is highly dependent on the optimizer. As we move to the second layer of $12$ qubit simulations on Fixed hamiltonian, we observe a decline in precision as the level of optimization rises. In this case, COBYLA winds up with $80\%$, L\_BGFS\_B with $70\%$, and SLSQP with $84\%$. Here, SLSQP's accuracy loss is less than that of the other two optimizers. The variable hamiltonian with $12$ qubits demonstrates a comparable trend. On the initial layer, we observe high accuracy with COBYLA at $96\%$, L\_BGFS\_B at $86\%$, and SLSQP at $90\%$. Moving to the second stratum, the accuracy figures drop significantly, with COBYLA at $76\%$ and L\_BGFS\_B at $62\%$, while SLSQP maintains excellent accuracy at $86\%$. In every scenario of our investigation, it is evident that over-optimization reduces accuracy rates.

\begin{figure*}[!ht]
\centering
\begin{subfigure}{0.5\linewidth}
\includegraphics[height=3.8cm,width=8cm]{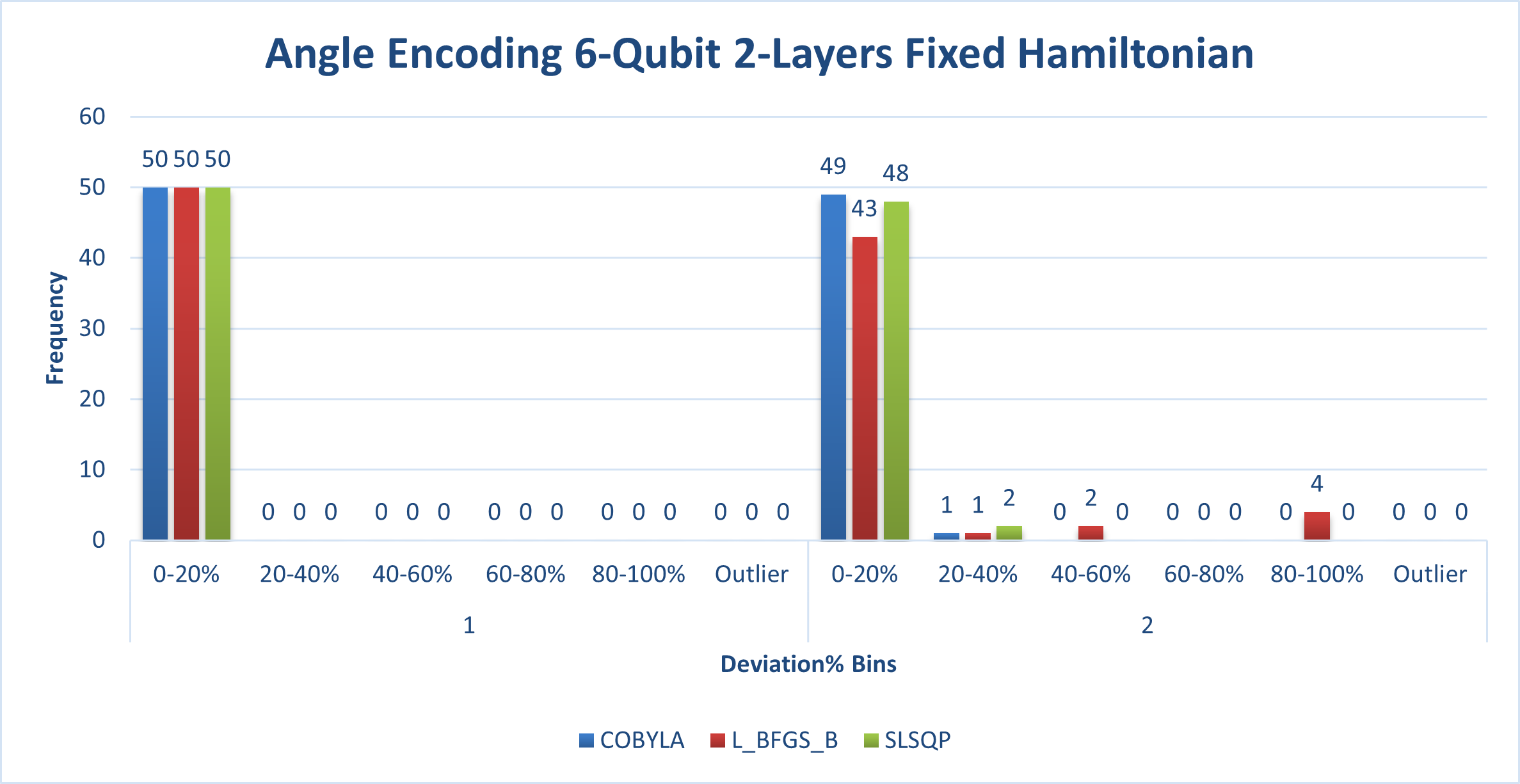} 
\caption{}
\label{}
\end{subfigure}\hfill
\begin{subfigure}{0.5\linewidth}
\includegraphics[height=3.8cm,width=8cm]{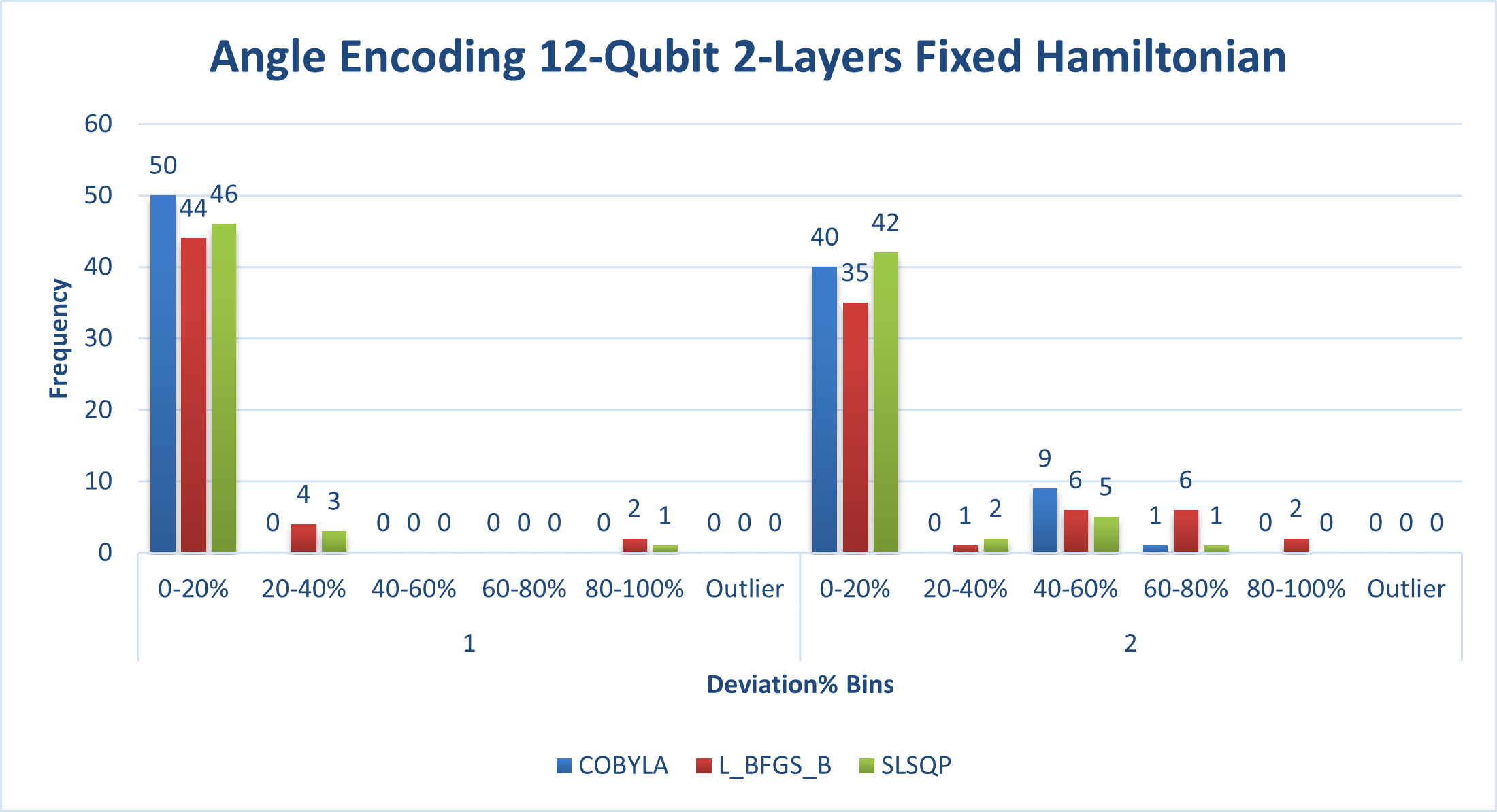} 
\caption{}
\label{}
\end{subfigure}\hfill
\begin{subfigure}{0.5\linewidth}
\includegraphics[height=3.8cm,width=8cm]{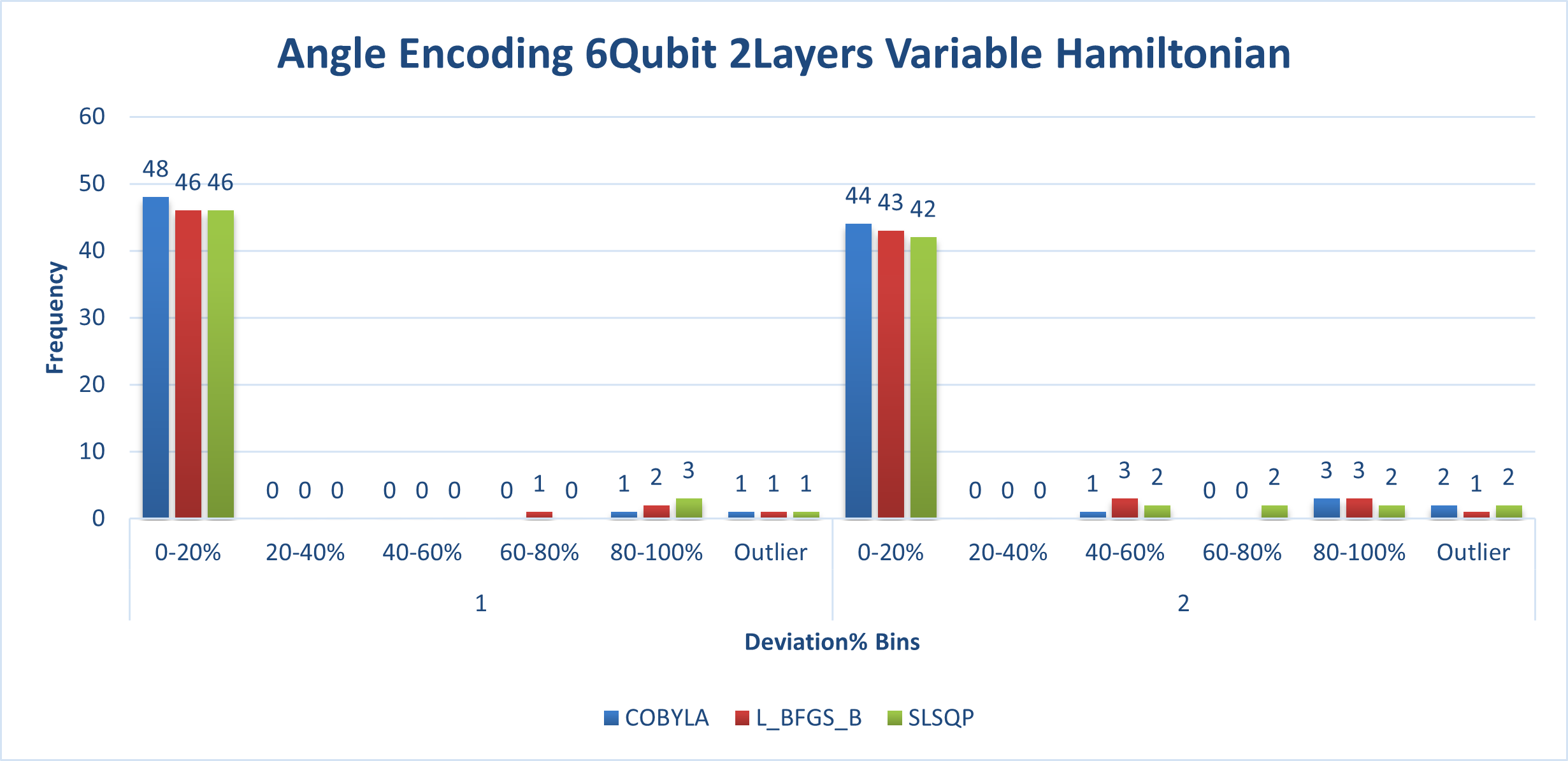} 
\caption{}
\label{}
\end{subfigure}\hfill
\begin{subfigure}{0.5\linewidth}
\includegraphics[height=3.8cm,width=8cm]{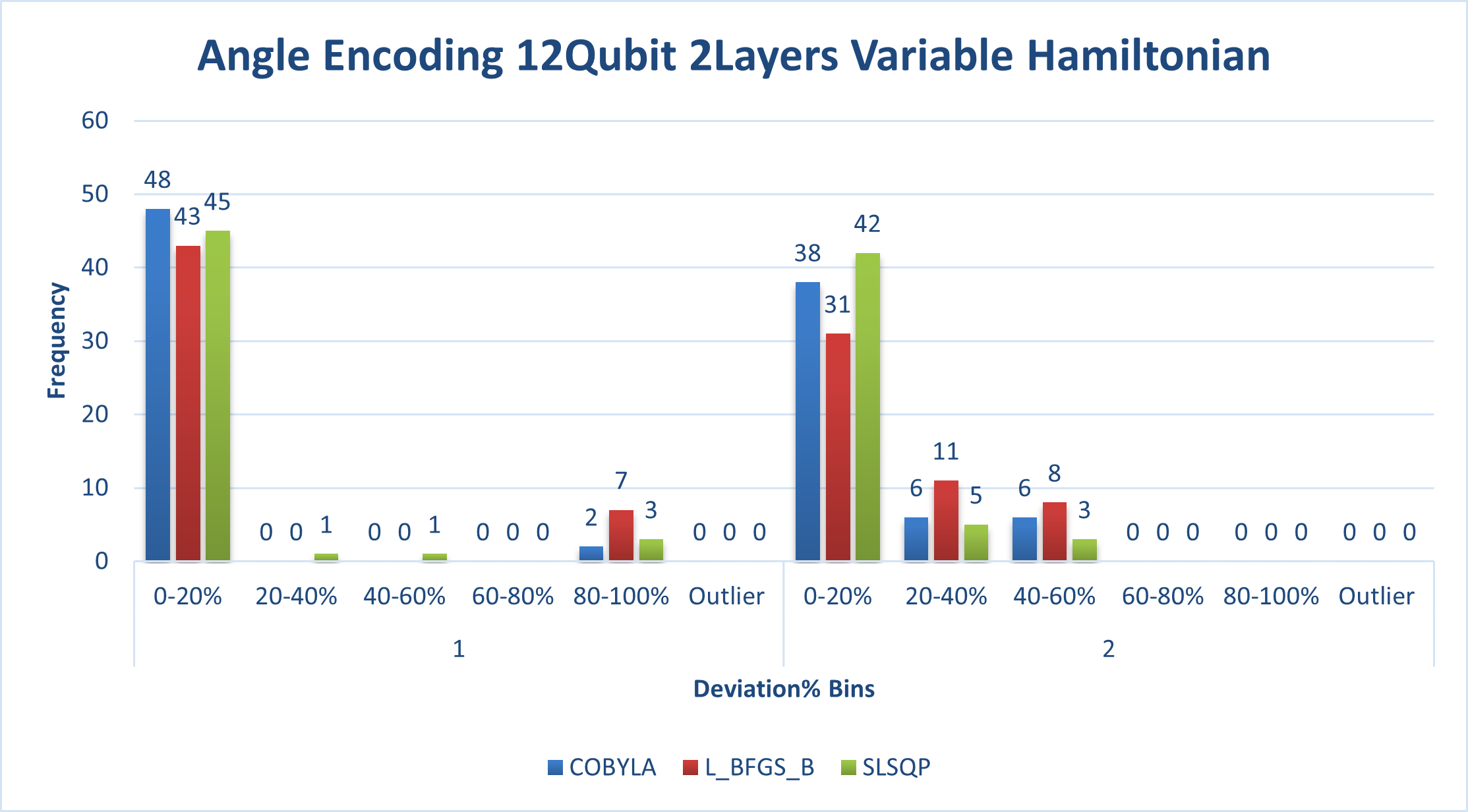} 
\caption{}
\label{}
\end{subfigure}\hfill
\caption{Plot illustrating angle encoding results for QSVM solution of VRP. (a) Angle encoding $6$ qubits Fix hamiltonian, (b) Angle encoding $12$ qubits Fix hamiltonian, (c) Angle encoding $6$ qubits Variable hamiltonian, (d) Angle encoding $12$ qubits Variable hamiltonian.}
\label{AngleEncodingResults}
\end{figure*}
\subsubsection{Higher Order Encoding}

After Amplitude and Angle Encoding, Higher Order Encoding is the third most prevalent encoding in our SVM VRP simulation experiment. This is also the third most accurate encoding in our experiment. For both $6$ qubit and $12$ qubit simulations, HO encoding yields moderately accurate results; however, as the number of circuit layers is increased, the accuracy of the HO encoding scheme deteriorates, rendering it inappropriate. Figure $5$ depicts the statistics of the HO encoding scheme for fixed and variable hamiltonian simulations of SVM VRP circuits over $50$ iterations for both $6$ qubit and $12$ qubit simulations. COBYLA achieves $78\%$ accuracy for a $6$-qubit HO encoding circuit on a fixed Hamiltonian, while L\_BGFS\_B achieves $66\%$ accuracy and SLSQP achieves $70\%$ accuracy. As we proceed to the second layer, the accuracy considerably decreases, with COBYLA at $34\%$ and SLSQP, L\_BGFS\_B at $16\%$, respectively. Similar trends can be observed in variable Hamiltonian simulations of HO encoding with $6$ qubits, with COBYLA at $76\%$, SLSQP at $62\%$, and L\_BGFS\_B at $58\%$ for the first layer; for the second layer, the accuracy drops to $36\%$, $34\%$, and $36\%$ for COBYLA, L\_BGFS\_B, and SLSQP, respectively. The $12$ qubit simulation yields superior results than the $6$ qubit simulation and improves COBYLA's accuracy. For fixed hamiltonian simulations, COBYLA achieves an accuracy of $92\%$, compared to $78\%$ for 6qubit. For variable hamiltonian simulations, COBYLA stores $76\%$ for $6$ qubit in the first layer, and $92\%$ for $12$ qubit in the first layer. The tendencies for L\_BGFS\_B and SLSQP are ambiguous for both cases (fixed and variable hamiltonian simulations); it is reassuring to conclude that an increase in layer decreases accuracy and that COBYLA outperforms the other two optimizers and ensures stable performance.

\begin{figure*}[!ht]
\centering
\begin{subfigure}{0.5\linewidth}
\includegraphics[height=3.8cm,width=8cm]{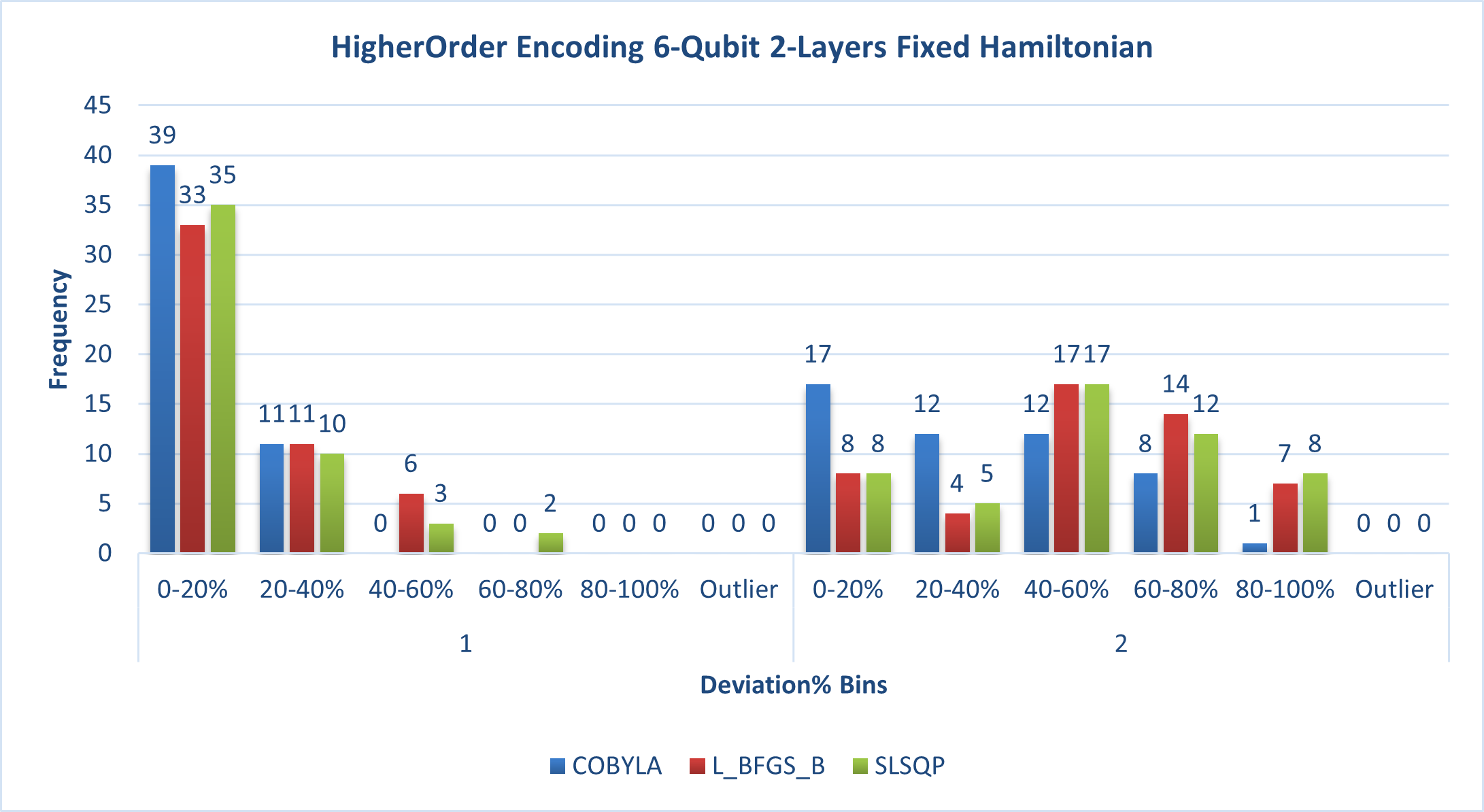} 
\caption{}
\label{}
\end{subfigure}\hfill
\begin{subfigure}{0.5\linewidth}
\includegraphics[height=3.8cm,width=8cm]{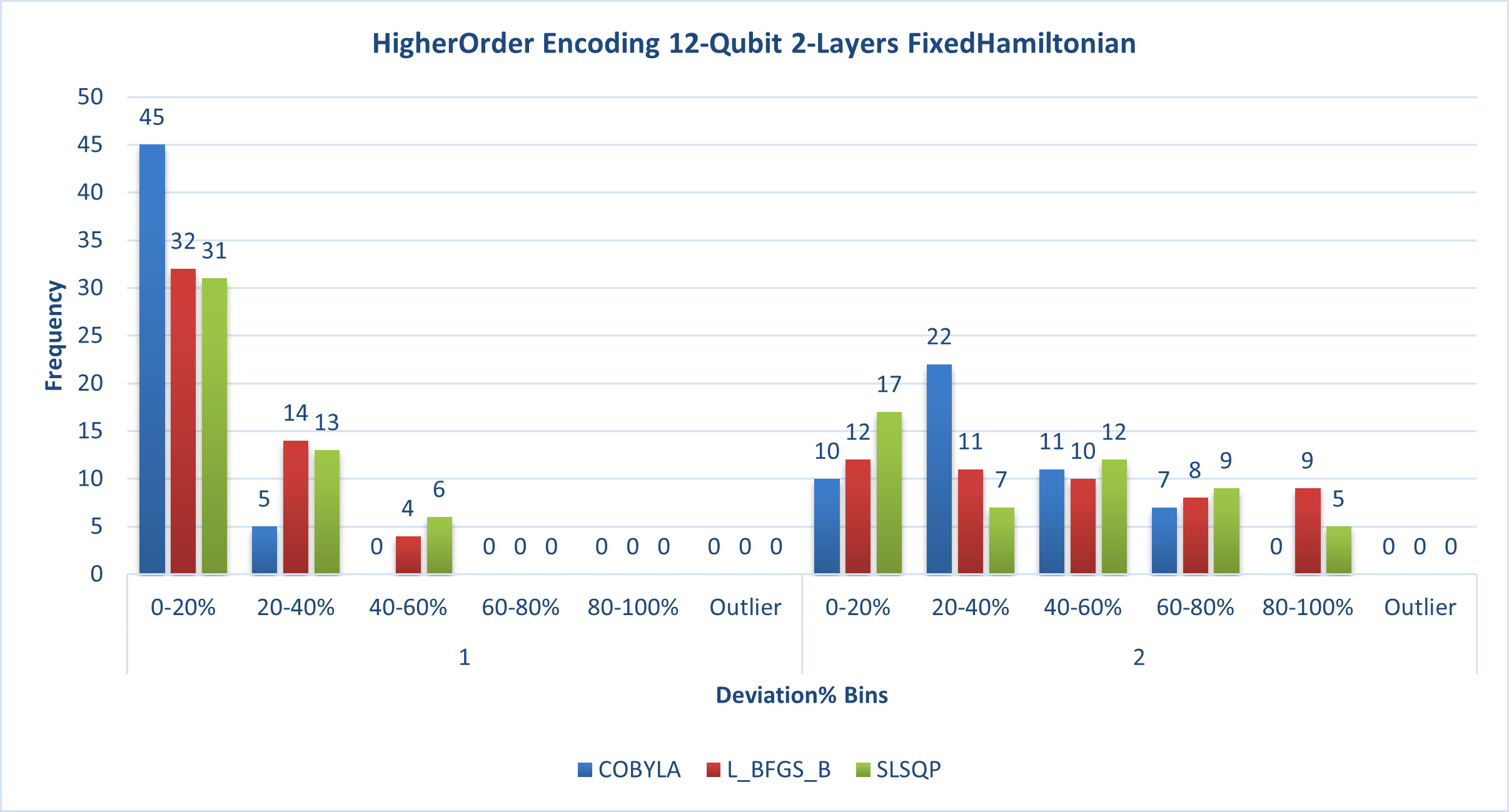} 
\caption{}
\label{}
\end{subfigure}\hfill
\begin{subfigure}{0.5\linewidth}
\includegraphics[height=3.8cm,width=8cm]{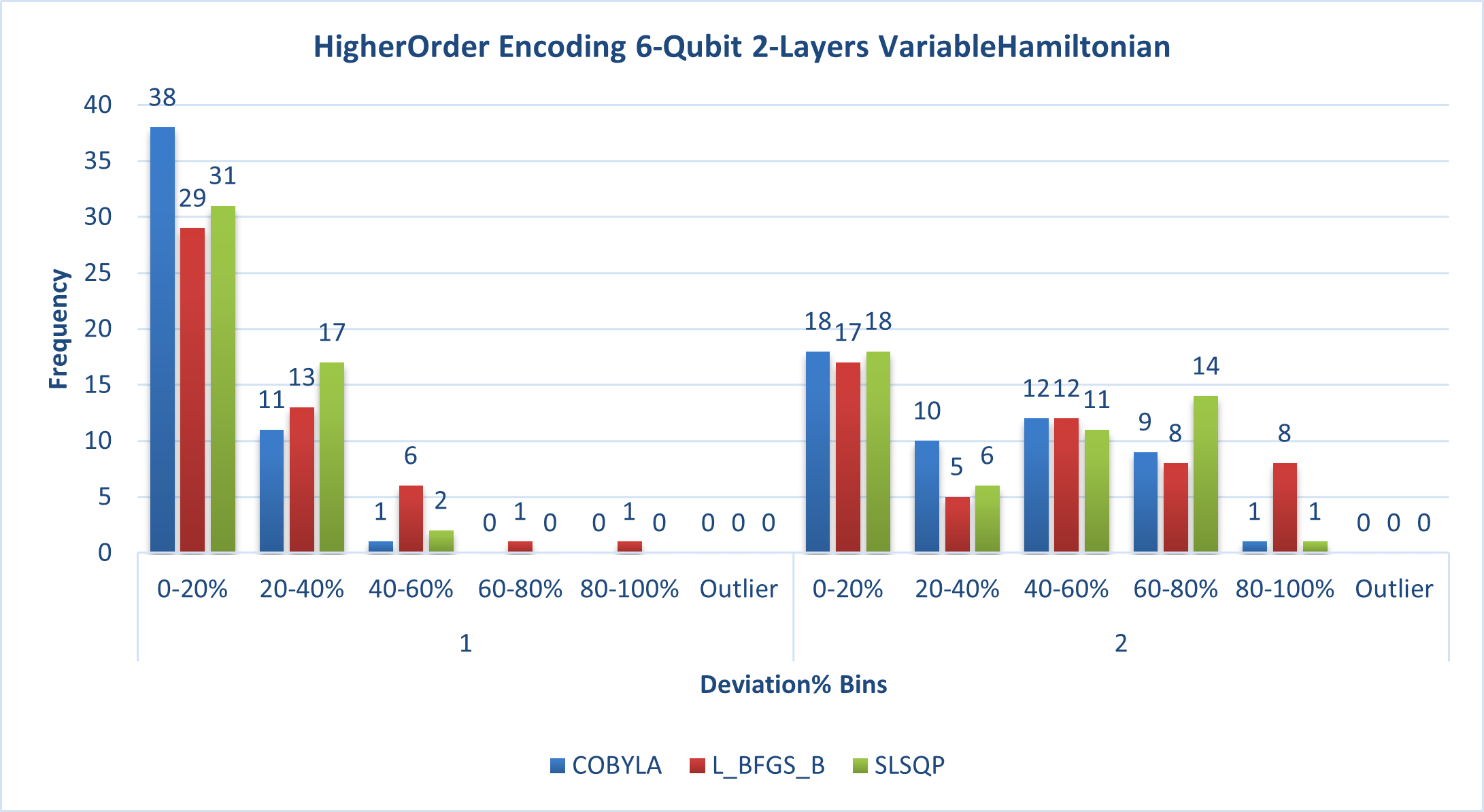} 
\caption{}
\label{}
\end{subfigure}\hfill
\begin{subfigure}{0.5\linewidth}
\includegraphics[height=3.8cm,width=8cm]{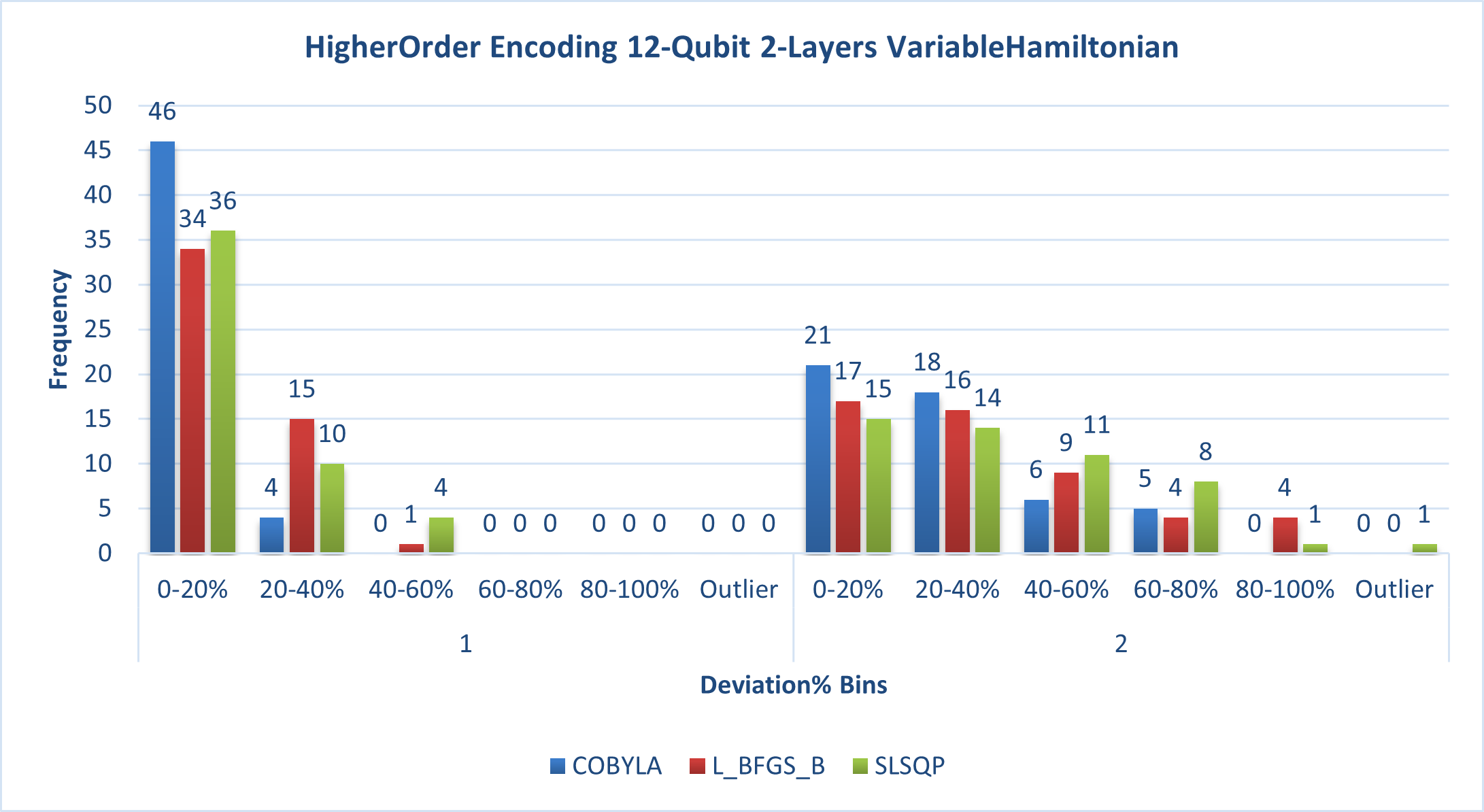} 
\caption{}
\label{}
\end{subfigure}\hfill
\caption{Plot illustrating Higherorder encoding results for QSVM solution of VRP. (a) Higherorder encoding $6$ qubits Fix hamiltonian, (b) Higherorder encoding $12$ qubits Fix hamiltonian, (c) Higherorder encoding $6$ qubits Variable hamiltonian, (d) Higherorder encoding $12$ qubits Variable hamiltonian.}
\label{HOEncodingResults}
\end{figure*}

\begin{figure*}[!ht]
\centering
\begin{subfigure}{0.5\linewidth}
\includegraphics[height=3.8cm,width=8cm]{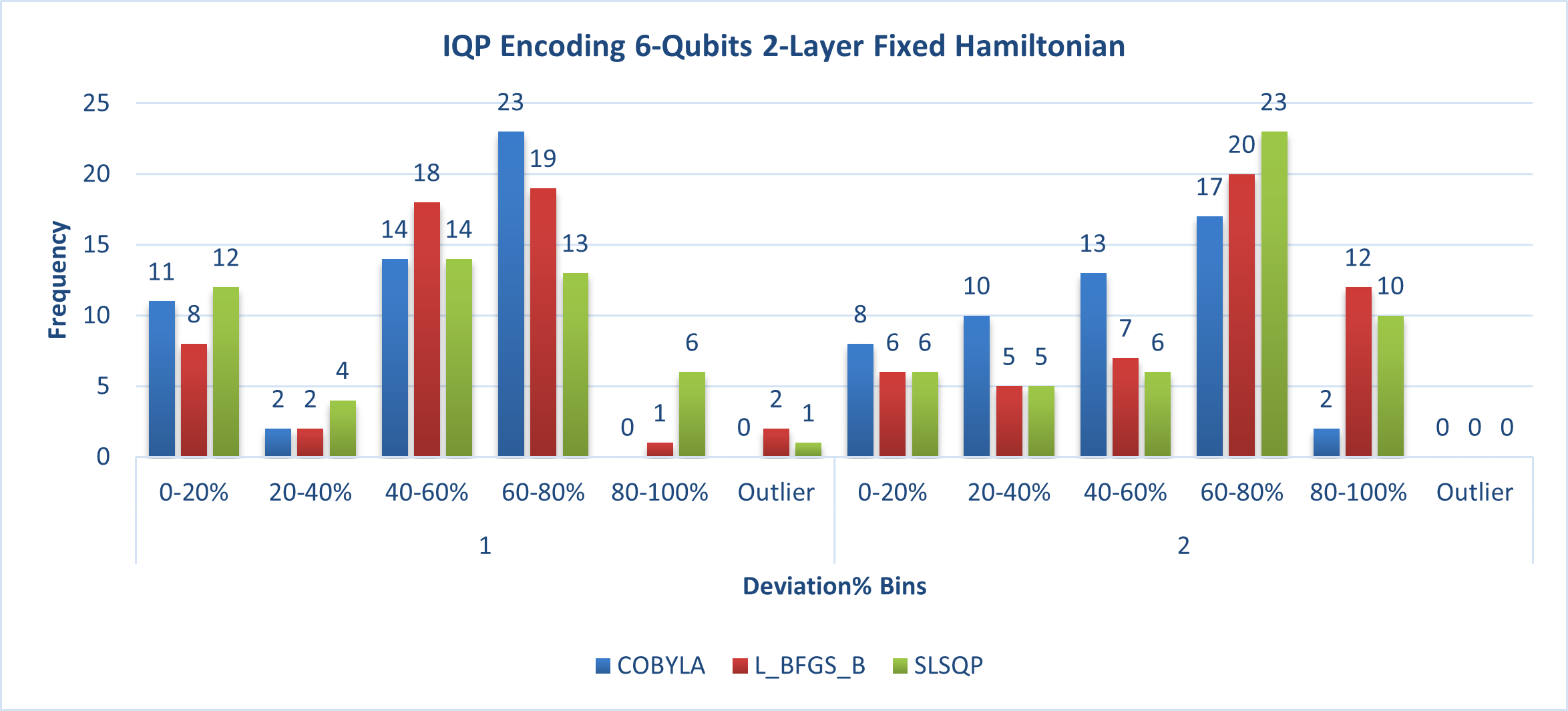} 
\caption{}
\label{}
\end{subfigure}\hfill
\begin{subfigure}{0.5\linewidth}
\includegraphics[height=3.8cm,width=8cm]{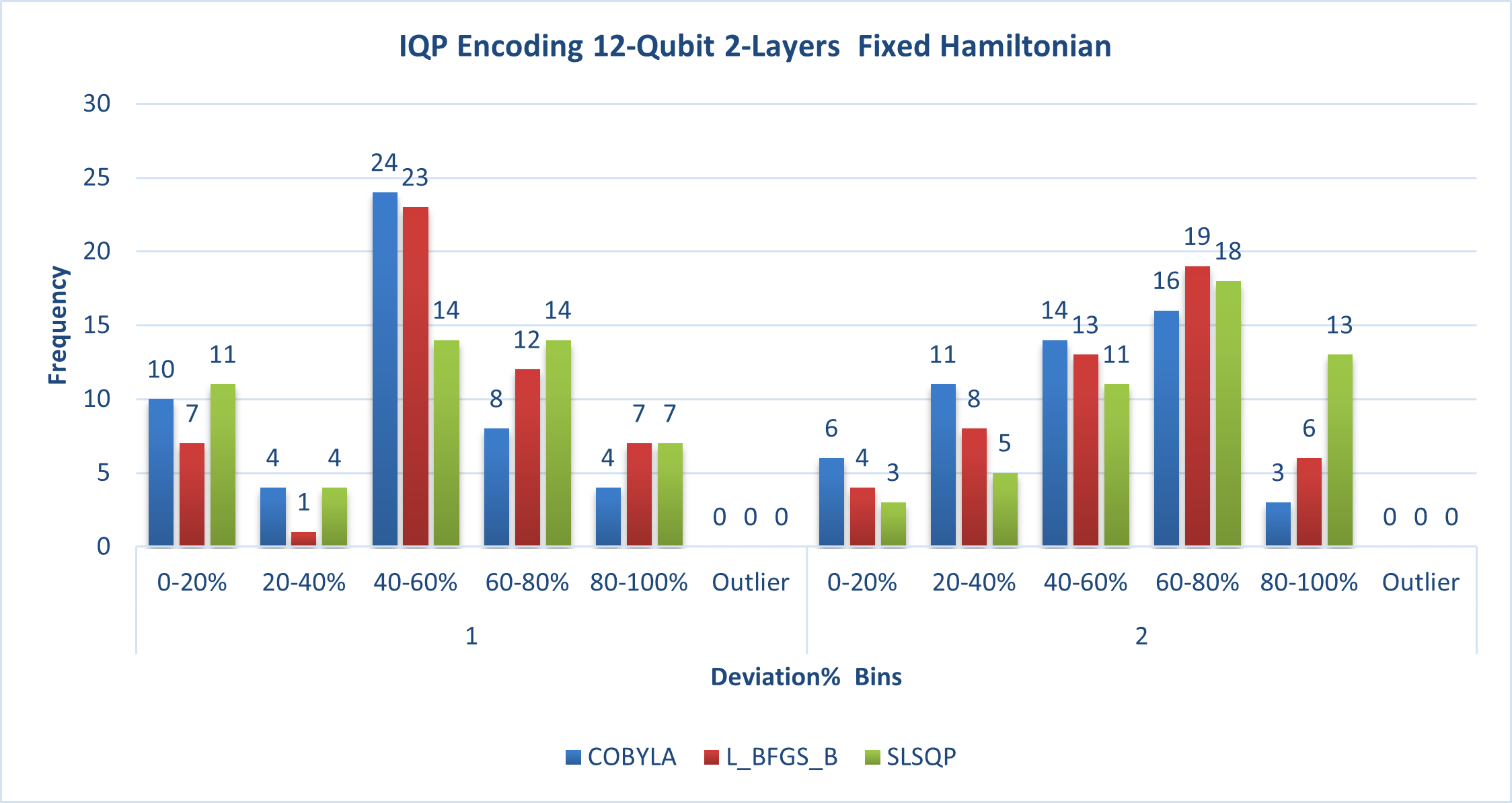} 
\caption{}
\label{}
\end{subfigure}\hfill
\begin{subfigure}{0.5\linewidth}
\includegraphics[height=3.8cm,width=8cm]{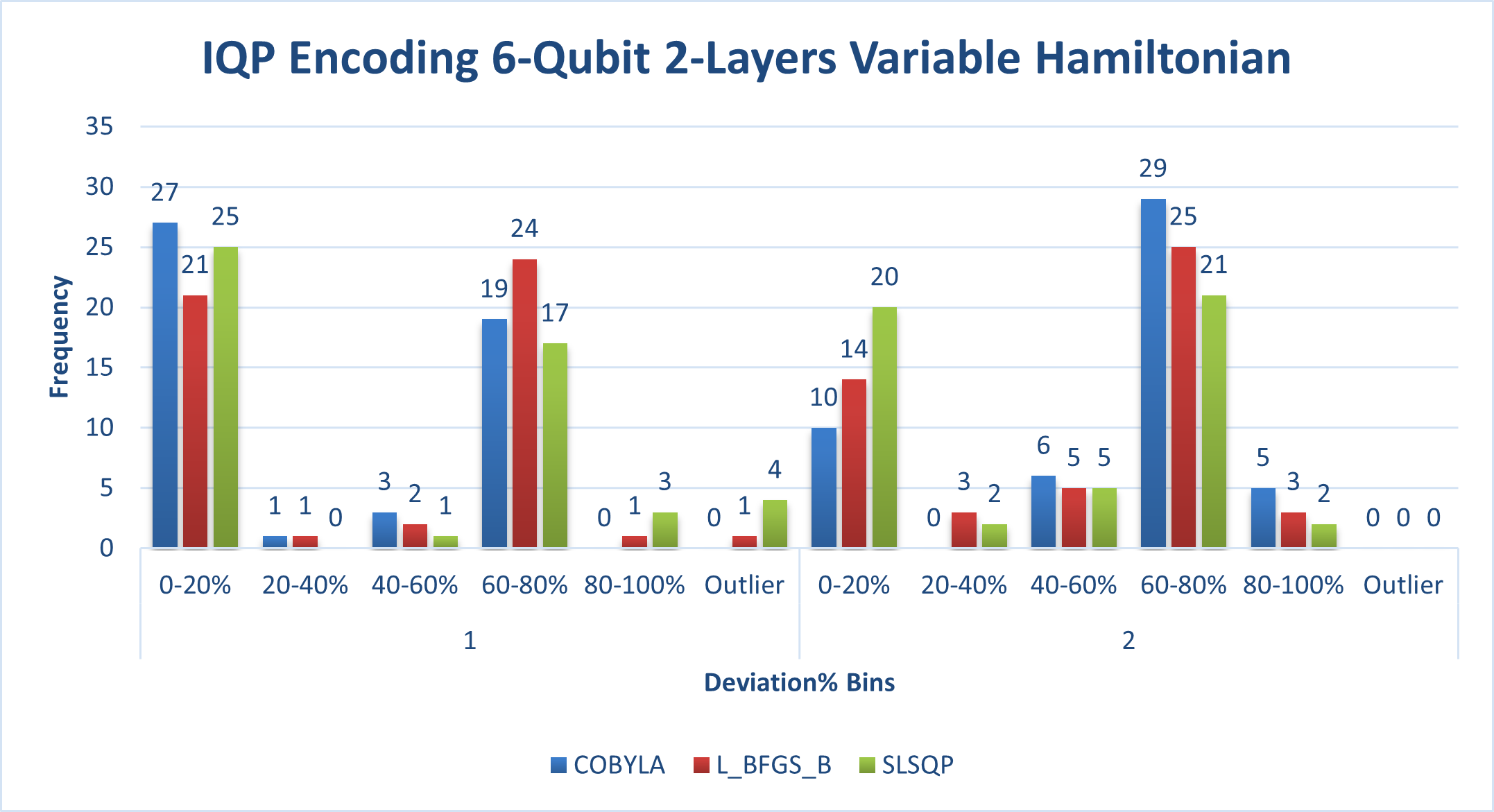}
\caption{}
\label{}
\end{subfigure}\hfill
\begin{subfigure}{0.5\linewidth}
\includegraphics[height=3.8cm,width=8cm]{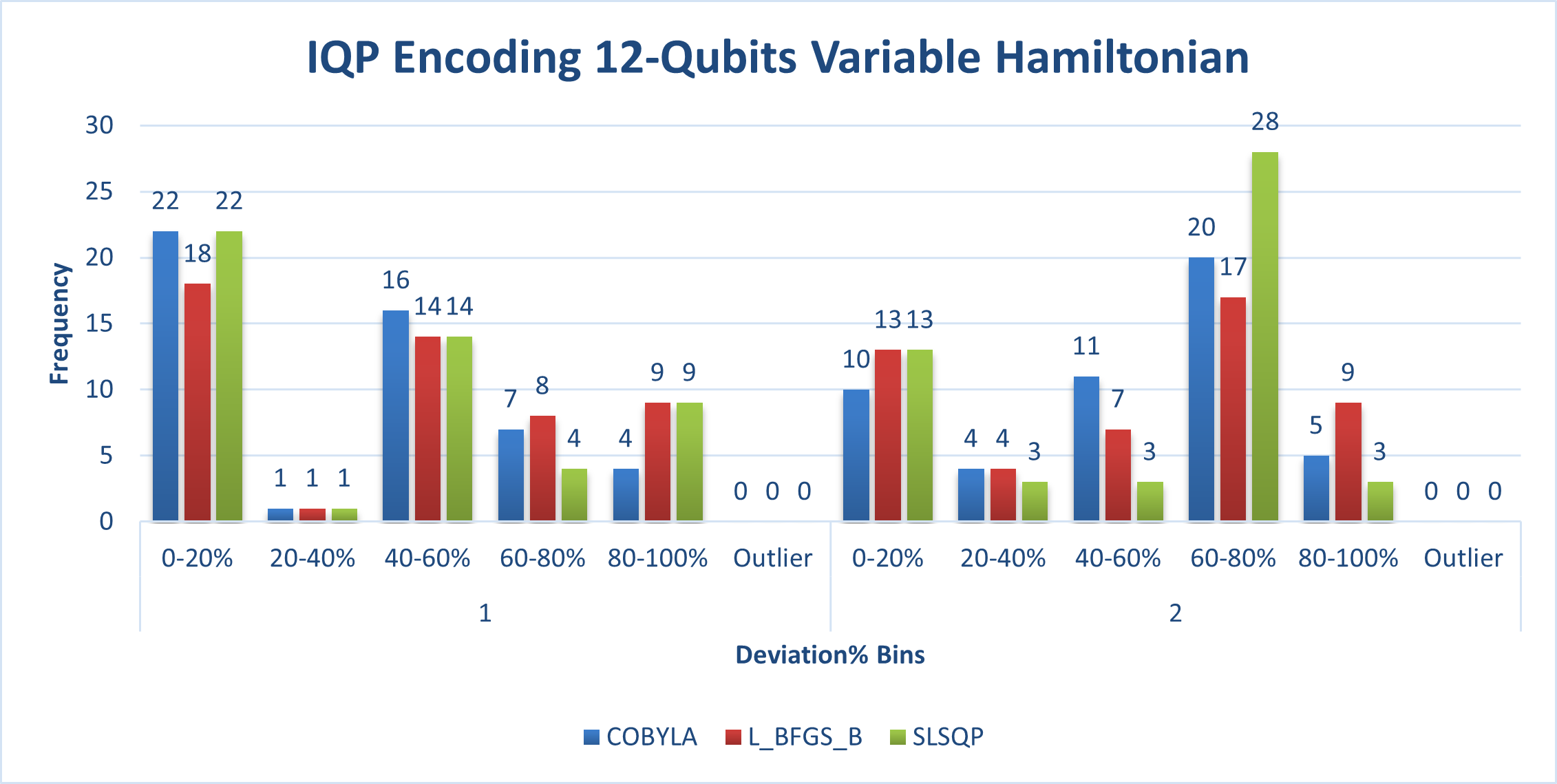} 
\caption{}
\label{}
\end{subfigure}\hfill
\caption{Plot illustrating IQP encoding results for QSVM solution of VRP. (a) IQP encoding $6$ qubits Fix hamiltonian, (b) IQP encoding $12$ qubits Fix hamiltonian, (c) IQP encoding $6$ qubits Variable hamiltonian, (d) IQP encoding $12$ qubits Variable hamiltonian.}
\label{IQPEncodingResults}
\end{figure*}

\subsubsection{IQP Encoding}

IQP encoding is the last and least accurate encoding in our experiment to simulate an SVM VRP circuit.  The results are plotted in \ref{IQPEncodingResults} and in tables \ref{Accuracy and Error All Encodings Fixed Hamiltonian} and tables \ref{Accuracy and Error All Encodings Variable hamiltonian}. As we can see from the figures and tables that accuracy is consistently poor for fixed and variable hamiltonian simulations in both $6$ qubit and $12$ qubit circuits. The accuracy further declines as layers increase. Hence this encoding is unsuitable in our experiment of SVM VRP circuits.

\begin{table*}[!htb]
\centering
\resizebox{1.5\columnwidth}{!}{
\arrayrulecolor{black}
\setlength{\extrarowheight}{0pt}
\addtolength{\extrarowheight}{\aboverulesep}
\addtolength{\extrarowheight}{\belowrulesep}
\setlength{\aboverulesep}{0pt}
\setlength{\belowrulesep}{0pt}
\begin{tabular}{@{}|cccc|cccc|cccc|@{}}
\toprule
\multicolumn{4}{|c|}{\textbf{Qubits}} &
  \multicolumn{4}{c|}{\textbf{6}} &
  \multicolumn{4}{c|}{\textbf{12}} \\ \midrule
\multicolumn{1}{|c|}{\textbf{Encoding}} &
  \multicolumn{1}{c|}{\textbf{Optimizer}} &
  \multicolumn{1}{c|}{\textbf{Layers}} &
  \textbf{Iterations} &
  \multicolumn{1}{c|}{\textbf{No Devn.}} &
  \multicolumn{1}{c|}{\textbf{With Devn}} &
  \multicolumn{1}{c|}{\textbf{Acc}} &
  \textbf{Err} &
  \multicolumn{1}{c|}{\textbf{No Devn.}} &
  \multicolumn{1}{c|}{\textbf{With Devn}} &
  \multicolumn{1}{c|}{\textbf{Acc}} &
  \textbf{Err} \\ \midrule
\multicolumn{1}{|c|}{\textbf{Amplitude Enc.}} &
  \multicolumn{1}{c|}{\textbf{COBYLA}} &
  \multicolumn{1}{c|}{1} &
  50 &
  \multicolumn{1}{c|}{\cellcolor[HTML]{63BE7B}50} &
  \multicolumn{1}{c|}{\cellcolor[HTML]{FCFCFF}0} &
  \multicolumn{1}{c|}{\cellcolor[HTML]{63BE7B}100\%} &
  \cellcolor[HTML]{FCFCFF}0\% &
  \multicolumn{1}{c|}{} &
  \multicolumn{1}{c|}{} &
  \multicolumn{1}{c|}{} &
   \\ \midrule
\multicolumn{1}{|c|}{\textbf{}} &
  \multicolumn{1}{c|}{\textbf{L\_BGFS\_B}} &
  \multicolumn{1}{c|}{1} &
  50 &
  \multicolumn{1}{c|}{\cellcolor[HTML]{63BE7B}50} &
  \multicolumn{1}{c|}{\cellcolor[HTML]{FCFCFF}0} &
  \multicolumn{1}{c|}{\cellcolor[HTML]{63BE7B}100\%} &
  \cellcolor[HTML]{FCFCFF}0\% &
  \multicolumn{1}{c|}{} &
  \multicolumn{1}{c|}{} &
  \multicolumn{1}{c|}{} &
   \\ \midrule
\multicolumn{1}{|c|}{\textbf{}} &
  \multicolumn{1}{c|}{\textbf{SLSQP}} &
  \multicolumn{1}{c|}{1} &
  50 &
  \multicolumn{1}{c|}{\cellcolor[HTML]{63BE7B}50} &
  \multicolumn{1}{c|}{\cellcolor[HTML]{FCFCFF}0} &
  \multicolumn{1}{c|}{\cellcolor[HTML]{63BE7B}100\%} &
  \cellcolor[HTML]{FCFCFF}0\% &
  \multicolumn{1}{c|}{} &
  \multicolumn{1}{c|}{} &
  \multicolumn{1}{c|}{} &
   \\ \midrule
\multicolumn{1}{|c|}{\textbf{}} &
  \multicolumn{1}{c|}{\textbf{COBYLA}} &
  \multicolumn{1}{c|}{2} &
  50 &
  \multicolumn{1}{c|}{\cellcolor[HTML]{63BE7B}50} &
  \multicolumn{1}{c|}{\cellcolor[HTML]{FCFCFF}0} &
  \multicolumn{1}{c|}{\cellcolor[HTML]{63BE7B}100\%} &
  \cellcolor[HTML]{FCFCFF}0\% &
  \multicolumn{1}{c|}{} &
  \multicolumn{1}{c|}{} &
  \multicolumn{1}{c|}{} &
   \\ \midrule
\multicolumn{1}{|c|}{\textbf{}} &
  \multicolumn{1}{c|}{\textbf{L\_BGFS\_B}} &
  \multicolumn{1}{c|}{2} &
  50 &
  \multicolumn{1}{c|}{\cellcolor[HTML]{63BE7B}50} &
  \multicolumn{1}{c|}{\cellcolor[HTML]{FCFCFF}0} &
  \multicolumn{1}{c|}{\cellcolor[HTML]{63BE7B}100\%} &
  \cellcolor[HTML]{FCFCFF}0\% &
  \multicolumn{1}{c|}{} &
  \multicolumn{1}{c|}{} &
  \multicolumn{1}{c|}{} &
   \\ \midrule
\multicolumn{1}{|c|}{\textbf{}} &
  \multicolumn{1}{c|}{\textbf{SLSQP}} &
  \multicolumn{1}{c|}{2} &
  50 &
  \multicolumn{1}{c|}{\cellcolor[HTML]{63BE7B}50} &
  \multicolumn{1}{c|}{\cellcolor[HTML]{FCFCFF}0} &
  \multicolumn{1}{c|}{\cellcolor[HTML]{63BE7B}100\%} &
  \cellcolor[HTML]{FCFCFF}0\% &
  \multicolumn{1}{c|}{} &
  \multicolumn{1}{c|}{} &
  \multicolumn{1}{c|}{} &
   \\ \midrule
\multicolumn{1}{|c|}{\textbf{Angle Enc.}} &
  \multicolumn{1}{c|}{\textbf{COBYLA}} &
  \multicolumn{1}{c|}{1} &
  50 &
  \multicolumn{1}{c|}{\cellcolor[HTML]{63BE7B}50} &
  \multicolumn{1}{c|}{\cellcolor[HTML]{FCFCFF}0} &
  \multicolumn{1}{c|}{\cellcolor[HTML]{63BE7B}100\%} &
  \cellcolor[HTML]{FCFCFF}0\% &
  \multicolumn{1}{c|}{\cellcolor[HTML]{00B0F0}50} &
  \multicolumn{1}{c|}{\cellcolor[HTML]{FFFFFF}0} &
  \multicolumn{1}{c|}{\cellcolor[HTML]{00B0F0}100\%} &
  \cellcolor[HTML]{FFFFFF}0\% \\ \midrule
\multicolumn{1}{|c|}{\textbf{}} &
  \multicolumn{1}{c|}{\textbf{L\_BGFS\_B}} &
  \multicolumn{1}{c|}{1} &
  50 &
  \multicolumn{1}{c|}{\cellcolor[HTML]{63BE7B}50} &
  \multicolumn{1}{c|}{\cellcolor[HTML]{FCFCFF}0} &
  \multicolumn{1}{c|}{\cellcolor[HTML]{63BE7B}100\%} &
  \cellcolor[HTML]{FCFCFF}0\% &
  \multicolumn{1}{c|}{\cellcolor[HTML]{22BBF2}44} &
  \multicolumn{1}{c|}{\cellcolor[HTML]{FEF4F5}6} &
  \multicolumn{1}{c|}{\cellcolor[HTML]{22BBF2}88\%} &
  \cellcolor[HTML]{FFF3F6}12\% \\ \midrule
\multicolumn{1}{|c|}{\textbf{}} &
  \multicolumn{1}{c|}{\textbf{SLSQP}} &
  \multicolumn{1}{c|}{1} &
  50 &
  \multicolumn{1}{c|}{\cellcolor[HTML]{63BE7B}50} &
  \multicolumn{1}{c|}{\cellcolor[HTML]{FCFCFF}0} &
  \multicolumn{1}{c|}{\cellcolor[HTML]{63BE7B}100\%} &
  \cellcolor[HTML]{FCFCFF}0\% &
  \multicolumn{1}{c|}{\cellcolor[HTML]{17B7F2}46} &
  \multicolumn{1}{c|}{\cellcolor[HTML]{FEF8F8}4} &
  \multicolumn{1}{c|}{\cellcolor[HTML]{17B7F2}92\%} &
  \cellcolor[HTML]{FFF7F9}8\% \\ \midrule
\multicolumn{1}{|c|}{\textbf{}} &
  \multicolumn{1}{c|}{\textbf{COBYLA}} &
  \multicolumn{1}{c|}{2} &
  50 &
  \multicolumn{1}{c|}{\cellcolor[HTML]{67C07E}49} &
  \multicolumn{1}{c|}{\cellcolor[HTML]{FCF9FC}1} &
  \multicolumn{1}{c|}{\cellcolor[HTML]{67C07E}98\%} &
  \cellcolor[HTML]{FCF9FC}2\% &
  \multicolumn{1}{c|}{\cellcolor[HTML]{38C2F4}40} &
  \multicolumn{1}{c|}{\cellcolor[HTML]{FDECED}10} &
  \multicolumn{1}{c|}{\cellcolor[HTML]{38C2F4}80\%} &
  \cellcolor[HTML]{FFEBEF}20\% \\ \midrule
\multicolumn{1}{|c|}{\textbf{}} &
  \multicolumn{1}{c|}{\textbf{L\_BGFS\_B}} &
  \multicolumn{1}{c|}{2} &
  50 &
  \multicolumn{1}{c|}{\cellcolor[HTML]{7CC890}43} &
  \multicolumn{1}{c|}{\cellcolor[HTML]{FCE5E8}7} &
  \multicolumn{1}{c|}{\cellcolor[HTML]{7CC890}86\%} &
  \cellcolor[HTML]{FCE5E8}14\% &
  \multicolumn{1}{c|}{\cellcolor[HTML]{54CAF5}35} &
  \multicolumn{1}{c|}{\cellcolor[HTML]{FCE2E4}15} &
  \multicolumn{1}{c|}{\cellcolor[HTML]{54CAF5}70\%} &
  \cellcolor[HTML]{FFE1E7}30\% \\ \midrule
\multicolumn{1}{|c|}{\textbf{}} &
  \multicolumn{1}{c|}{\textbf{SLSQP}} &
  \multicolumn{1}{c|}{2} &
  50 &
  \multicolumn{1}{c|}{\cellcolor[HTML]{6AC181}48} &
  \multicolumn{1}{c|}{\cellcolor[HTML]{FCF6F9}2} &
  \multicolumn{1}{c|}{\cellcolor[HTML]{6AC182}96\%} &
  \cellcolor[HTML]{FCF6F9}4\% &
  \multicolumn{1}{c|}{\cellcolor[HTML]{2DBEF3}42} &
  \multicolumn{1}{c|}{\cellcolor[HTML]{FDF0F1}8} &
  \multicolumn{1}{c|}{\cellcolor[HTML]{2DBEF3}84\%} &
  \cellcolor[HTML]{FFEFF2}16\% \\ \midrule
\multicolumn{1}{|c|}{\textbf{HO Enc.}} &
  \multicolumn{1}{c|}{\textbf{COBYLA}} &
  \multicolumn{1}{c|}{1} &
  50 &
  \multicolumn{1}{c|}{\cellcolor[HTML]{8ACE9C}39} &
  \multicolumn{1}{c|}{\cellcolor[HTML]{FBD8DA}11} &
  \multicolumn{1}{c|}{\cellcolor[HTML]{8ACE9C}78\%} &
  \cellcolor[HTML]{FBD8DA}22\% &
  \multicolumn{1}{c|}{\cellcolor[HTML]{1CB9F2}45} &
  \multicolumn{1}{c|}{\cellcolor[HTML]{FEF6F6}5} &
  \multicolumn{1}{c|}{\cellcolor[HTML]{1CB9F2}90\%} &
  \cellcolor[HTML]{FFF5F7}10\% \\ \midrule
\multicolumn{1}{|c|}{\textbf{}} &
  \multicolumn{1}{c|}{\textbf{L\_BGFS\_B}} &
  \multicolumn{1}{c|}{1} &
  50 &
  \multicolumn{1}{c|}{\cellcolor[HTML]{9FD6AE}33} &
  \multicolumn{1}{c|}{\cellcolor[HTML]{FBC4C6}17} &
  \multicolumn{1}{c|}{\cellcolor[HTML]{9FD6AE}66\%} &
  \cellcolor[HTML]{FBC4C6}34\% &
  \multicolumn{1}{c|}{\cellcolor[HTML]{64CFF6}32} &
  \multicolumn{1}{c|}{\cellcolor[HTML]{FBDDDF}18} &
  \multicolumn{1}{c|}{\cellcolor[HTML]{64CFF6}64\%} &
  \cellcolor[HTML]{FEDAE2}36\% \\ \midrule
\multicolumn{1}{|c|}{\textbf{}} &
  \multicolumn{1}{c|}{\textbf{SLSQP}} &
  \multicolumn{1}{c|}{1} &
  50 &
  \multicolumn{1}{c|}{\cellcolor[HTML]{98D4A8}35} &
  \multicolumn{1}{c|}{\cellcolor[HTML]{FBCACD}15} &
  \multicolumn{1}{c|}{\cellcolor[HTML]{98D4A8}70\%} &
  \cellcolor[HTML]{FBCACD}30\% &
  \multicolumn{1}{c|}{\cellcolor[HTML]{6AD1F7}31} &
  \multicolumn{1}{c|}{\cellcolor[HTML]{FADBDD}19} &
  \multicolumn{1}{c|}{\cellcolor[HTML]{6AD1F7}62\%} &
  \cellcolor[HTML]{FED8E0}38\% \\ \midrule
\multicolumn{1}{|c|}{\textbf{}} &
  \multicolumn{1}{c|}{\textbf{COBYLA}} &
  \multicolumn{1}{c|}{2} &
  50 &
  \multicolumn{1}{c|}{\cellcolor[HTML]{D6EDDE}17} &
  \multicolumn{1}{c|}{\cellcolor[HTML]{F98E90}33} &
  \multicolumn{1}{c|}{\cellcolor[HTML]{D6EDDE}34\%} &
  \cellcolor[HTML]{F98E90}66\% &
  \multicolumn{1}{c|}{\cellcolor[HTML]{DEF5FE}10} &
  \multicolumn{1}{c|}{\cellcolor[HTML]{F5B2B7}40} &
  \multicolumn{1}{c|}{\cellcolor[HTML]{DEF5FE}20\%} &
  \cellcolor[HTML]{FDADBE}80\% \\ \midrule
\multicolumn{1}{|c|}{\textbf{}} &
  \multicolumn{1}{c|}{\textbf{L\_BGFS\_B}} &
  \multicolumn{1}{c|}{2} &
  50 &
  \multicolumn{1}{c|}{\cellcolor[HTML]{F6FAF9}8} &
  \multicolumn{1}{c|}{\cellcolor[HTML]{F97072}42} &
  \multicolumn{1}{c|}{\cellcolor[HTML]{F6FAF9}16\%} &
  \cellcolor[HTML]{F97072}84\% &
  \multicolumn{1}{c|}{\cellcolor[HTML]{D3F2FD}12} &
  \multicolumn{1}{c|}{\cellcolor[HTML]{F5B6BA}38} &
  \multicolumn{1}{c|}{\cellcolor[HTML]{D3F2FD}24\%} &
  \cellcolor[HTML]{FDB1C1}76\% \\ \midrule
\multicolumn{1}{|c|}{\textbf{}} &
  \multicolumn{1}{c|}{\textbf{SLSQP}} &
  \multicolumn{1}{c|}{2} &
  50 &
  \multicolumn{1}{c|}{\cellcolor[HTML]{F6FAF9}8} &
  \multicolumn{1}{c|}{\cellcolor[HTML]{F97072}42} &
  \multicolumn{1}{c|}{\cellcolor[HTML]{F6FAF9}16\%} &
  \cellcolor[HTML]{F97072}84\% &
  \multicolumn{1}{c|}{\cellcolor[HTML]{B7E9FB}17} &
  \multicolumn{1}{c|}{\cellcolor[HTML]{F7C0C4}33} &
  \multicolumn{1}{c|}{\cellcolor[HTML]{B7E9FB}34\%} &
  \cellcolor[HTML]{FDBBCA}66\% \\ \midrule
\multicolumn{1}{|c|}{\textbf{IQP Enc.}} &
  \multicolumn{1}{c|}{\textbf{COBYLA}} &
  \multicolumn{1}{c|}{1} &
  50 &
  \multicolumn{1}{c|}{\cellcolor[HTML]{EBF5F0}11} &
  \multicolumn{1}{c|}{\cellcolor[HTML]{F97A7C}39} &
  \multicolumn{1}{c|}{\cellcolor[HTML]{EBF5F0}22\%} &
  \cellcolor[HTML]{F97A7C}78\% &
  \multicolumn{1}{c|}{\cellcolor[HTML]{DEF5FE}10} &
  \multicolumn{1}{c|}{\cellcolor[HTML]{F5B2B7}40} &
  \multicolumn{1}{c|}{\cellcolor[HTML]{DEF5FE}20\%} &
  \cellcolor[HTML]{FDADBE}80\% \\ \midrule
\multicolumn{1}{|c|}{\textbf{}} &
  \multicolumn{1}{c|}{\textbf{L\_BGFS\_B}} &
  \multicolumn{1}{c|}{1} &
  50 &
  \multicolumn{1}{c|}{\cellcolor[HTML]{F6FAF9}8} &
  \multicolumn{1}{c|}{\cellcolor[HTML]{F97072}42} &
  \multicolumn{1}{c|}{\cellcolor[HTML]{F6FAF9}16\%} &
  \cellcolor[HTML]{F97072}84\% &
  \multicolumn{1}{c|}{\cellcolor[HTML]{EFFAFF}7} &
  \multicolumn{1}{c|}{\cellcolor[HTML]{F4ACB1}43} &
  \multicolumn{1}{c|}{\cellcolor[HTML]{EFFAFF}14\%} &
  \cellcolor[HTML]{FDA7B9}86\% \\ \midrule
\multicolumn{1}{|c|}{\textbf{}} &
  \multicolumn{1}{c|}{\textbf{SLSQP}} &
  \multicolumn{1}{c|}{1} &
  50 &
  \multicolumn{1}{c|}{\cellcolor[HTML]{E8F4ED}12} &
  \multicolumn{1}{c|}{\cellcolor[HTML]{F97E80}38} &
  \multicolumn{1}{c|}{\cellcolor[HTML]{E8F4ED}24\%} &
  \cellcolor[HTML]{F97E80}76\% &
  \multicolumn{1}{c|}{\cellcolor[HTML]{D9F3FD}11} &
  \multicolumn{1}{c|}{\cellcolor[HTML]{F2A0A6}49} &
  \multicolumn{1}{c|}{\cellcolor[HTML]{D9F3FD}22\%} &
  \cellcolor[HTML]{FC9AAF}98\% \\ \midrule
\multicolumn{1}{|c|}{\textbf{}} &
  \multicolumn{1}{c|}{\textbf{COBYLA}} &
  \multicolumn{1}{c|}{2} &
  50 &
  \multicolumn{1}{c|}{\cellcolor[HTML]{F6FAF9}8} &
  \multicolumn{1}{c|}{\cellcolor[HTML]{F97072}42} &
  \multicolumn{1}{c|}{\cellcolor[HTML]{F6FAF9}16\%} &
  \cellcolor[HTML]{F97072}84\% &
  \multicolumn{1}{c|}{\cellcolor[HTML]{F4FCFF}6} &
  \multicolumn{1}{c|}{\cellcolor[HTML]{F4AAB0}44} &
  \multicolumn{1}{c|}{\cellcolor[HTML]{F4FCFF}12\%} &
  \cellcolor[HTML]{FDA5B8}88\% \\ \midrule
\multicolumn{1}{|c|}{\textbf{}} &
  \multicolumn{1}{c|}{\textbf{L\_BGFS\_B}} &
  \multicolumn{1}{c|}{2} &
  50 &
  \multicolumn{1}{c|}{\cellcolor[HTML]{FCFCFF}6} &
  \multicolumn{1}{c|}{\cellcolor[HTML]{F8696B}44} &
  \multicolumn{1}{c|}{\cellcolor[HTML]{FCFCFF}12\%} &
  \cellcolor[HTML]{F8696B}88\% &
  \multicolumn{1}{c|}{\cellcolor[HTML]{FFFFFF}4} &
  \multicolumn{1}{c|}{\cellcolor[HTML]{F3A6AC}46} &
  \multicolumn{1}{c|}{\cellcolor[HTML]{FFFFFF}8\%} &
  \cellcolor[HTML]{FDA1B4}92\% \\ \midrule
\multicolumn{1}{|c|}{\textbf{}} &
  \multicolumn{1}{c|}{\textbf{SLSQP}} &
  \multicolumn{1}{c|}{2} &
  50 &
  \multicolumn{1}{c|}{\cellcolor[HTML]{FCFCFF}6} &
  \multicolumn{1}{c|}{\cellcolor[HTML]{F8696B}44} &
  \multicolumn{1}{c|}{\cellcolor[HTML]{FCFCFF}12\%} &
  \cellcolor[HTML]{F8696B}88\% &
  \multicolumn{1}{c|}{\cellcolor[HTML]{FFFFFF}4} &
  \multicolumn{1}{c|}{\cellcolor[HTML]{F3A6AC}46} &
  \multicolumn{1}{c|}{\cellcolor[HTML]{FFFFFF}8\%} &
  \cellcolor[HTML]{FDA1B4}92\% \\ \bottomrule
\end{tabular}}
\caption{For $6$ and $12$ qubit VRP circuits using SVM with $2$ layers , the table above shows the  Accuracy and Error with reference to classical minimum (over $50$ iterations) for VQE simulations over a fixed Hamiltonian; utilizing Amplitude, Angle, Higher-Order, and IQP encoding schemes, Over the use of COBYLA, SLSQP and L\_BGFS\_B optimizers.}
\label{Accuracy and Error All Encodings Fixed Hamiltonian}
\end{table*}

\begin{table*}[!htb]
\centering
\resizebox{1.5\columnwidth}{!}{
\setlength{\extrarowheight}{0pt}
\addtolength{\extrarowheight}{\aboverulesep}
\addtolength{\extrarowheight}{\belowrulesep}
\setlength{\aboverulesep}{0pt}
\setlength{\belowrulesep}{0pt}
\arrayrulecolor{black}
\begin{tabular}{@{}|cccc|cccc|cccc|@{}}
\toprule
\multicolumn{4}{|c|}{\textbf{Qubits}} &
  \multicolumn{4}{c|}{\textbf{6}} &
  \multicolumn{4}{c|}{\textbf{12}} \\ \midrule
\multicolumn{1}{|c|}{\textbf{Encoding}} &
  \multicolumn{1}{c|}{\textbf{Optimizer}} &
  \multicolumn{1}{c|}{\textbf{Layers}} &
  \textbf{Iterations} &
  \multicolumn{1}{c|}{\textbf{No Devn.}} &
  \multicolumn{1}{c|}{\textbf{With Devn}} &
  \multicolumn{1}{c|}{\textbf{Acc}} &
  \textbf{Err} &
  \multicolumn{1}{c|}{\textbf{No Devn.}} &
  \multicolumn{1}{c|}{\textbf{With Devn}} &
  \multicolumn{1}{c|}{\textbf{Acc}} &
  \textbf{Err} \\ \midrule
\multicolumn{1}{|c|}{\textbf{Amplitude Enc.}} &
  \multicolumn{1}{c|}{\textbf{COBYLA}} &
  \multicolumn{1}{c|}{1} &
  50 &
  \multicolumn{1}{c|}{\cellcolor[HTML]{63BE7B}48} &
  \multicolumn{1}{c|}{\cellcolor[HTML]{FCFCFF}2} &
  \multicolumn{1}{c|}{\cellcolor[HTML]{63BE7B}96\%} &
  \cellcolor[HTML]{FCFCFF}4\% &
  \multicolumn{1}{c|}{} &
  \multicolumn{1}{c|}{} &
  \multicolumn{1}{c|}{} &
   \\ \midrule
\multicolumn{1}{|c|}{\textbf{}} &
  \multicolumn{1}{c|}{\textbf{L\_BGFS\_B}} &
  \multicolumn{1}{c|}{1} &
  50 &
  \multicolumn{1}{c|}{\cellcolor[HTML]{63BE7B}48} &
  \multicolumn{1}{c|}{\cellcolor[HTML]{FCFCFF}2} &
  \multicolumn{1}{c|}{\cellcolor[HTML]{63BE7B}96\%} &
  \cellcolor[HTML]{FCFCFF}4\% &
  \multicolumn{1}{c|}{} &
  \multicolumn{1}{c|}{} &
  \multicolumn{1}{c|}{} &
   \\ \midrule
\multicolumn{1}{|c|}{\textbf{}} &
  \multicolumn{1}{c|}{\textbf{SLSQP}} &
  \multicolumn{1}{c|}{1} &
  50 &
  \multicolumn{1}{c|}{\cellcolor[HTML]{63BE7B}48} &
  \multicolumn{1}{c|}{\cellcolor[HTML]{FCFCFF}2} &
  \multicolumn{1}{c|}{\cellcolor[HTML]{63BE7B}96\%} &
  \cellcolor[HTML]{FCFCFF}4\% &
  \multicolumn{1}{c|}{} &
  \multicolumn{1}{c|}{} &
  \multicolumn{1}{c|}{} &
   \\ \midrule
\multicolumn{1}{|c|}{\textbf{}} &
  \multicolumn{1}{c|}{\textbf{COBYLA}} &
  \multicolumn{1}{c|}{2} &
  50 &
  \multicolumn{1}{c|}{\cellcolor[HTML]{68C07F}47} &
  \multicolumn{1}{c|}{\cellcolor[HTML]{FCF9FC}3} &
  \multicolumn{1}{c|}{\cellcolor[HTML]{68C07F}94\%} &
  \cellcolor[HTML]{FCF9FC}6\% &
  \multicolumn{1}{c|}{} &
  \multicolumn{1}{c|}{} &
  \multicolumn{1}{c|}{} &
   \\ \midrule
\multicolumn{1}{|c|}{\textbf{}} &
  \multicolumn{1}{c|}{\textbf{L\_BGFS\_B}} &
  \multicolumn{1}{c|}{2} &
  50 &
  \multicolumn{1}{c|}{\cellcolor[HTML]{68C07F}47} &
  \multicolumn{1}{c|}{\cellcolor[HTML]{FCF9FC}3} &
  \multicolumn{1}{c|}{\cellcolor[HTML]{68C07F}94\%} &
  \cellcolor[HTML]{FCF9FC}6\% &
  \multicolumn{1}{c|}{} &
  \multicolumn{1}{c|}{} &
  \multicolumn{1}{c|}{} &
   \\ \midrule
\multicolumn{1}{|c|}{\textbf{}} &
  \multicolumn{1}{c|}{\textbf{SLSQP}} &
  \multicolumn{1}{c|}{2} &
  50 &
  \multicolumn{1}{c|}{\cellcolor[HTML]{68C07F}47} &
  \multicolumn{1}{c|}{\cellcolor[HTML]{FCF9FC}3} &
  \multicolumn{1}{c|}{\cellcolor[HTML]{68C07F}94\%} &
  \cellcolor[HTML]{FCF9FC}6\% &
  \multicolumn{1}{c|}{} &
  \multicolumn{1}{c|}{} &
  \multicolumn{1}{c|}{} &
   \\ \midrule
\multicolumn{1}{|c|}{\textbf{Angle Enc.}} &
  \multicolumn{1}{c|}{\textbf{COBYLA}} &
  \multicolumn{1}{c|}{1} &
  50 &
  \multicolumn{1}{c|}{\cellcolor[HTML]{63BE7B}48} &
  \multicolumn{1}{c|}{\cellcolor[HTML]{FCFCFF}2} &
  \multicolumn{1}{c|}{\cellcolor[HTML]{63BE7B}96\%} &
  \cellcolor[HTML]{FCFCFF}4\% &
  \multicolumn{1}{c|}{\cellcolor[HTML]{00B0F0}48} &
  \multicolumn{1}{c|}{\cellcolor[HTML]{FFFFFF}2} &
  \multicolumn{1}{c|}{\cellcolor[HTML]{00B0F0}96\%} &
  \cellcolor[HTML]{FFFFFF}4\% \\ \midrule
\multicolumn{1}{|c|}{\textbf{}} &
  \multicolumn{1}{c|}{\textbf{L\_BGFS\_B}} &
  \multicolumn{1}{c|}{1} &
  50 &
  \multicolumn{1}{c|}{\cellcolor[HTML]{6CC282}46} &
  \multicolumn{1}{c|}{\cellcolor[HTML]{FCF5F8}4} &
  \multicolumn{1}{c|}{\cellcolor[HTML]{6CC282}92\%} &
  \cellcolor[HTML]{FCF5F8}8\% &
  \multicolumn{1}{c|}{\cellcolor[HTML]{22BBF2}43} &
  \multicolumn{1}{c|}{\cellcolor[HTML]{FEF3F4}7} &
  \multicolumn{1}{c|}{\cellcolor[HTML]{22BBF2}86\%} &
  \cellcolor[HTML]{FFF2F5}14\% \\ \midrule
\multicolumn{1}{|c|}{\textbf{}} &
  \multicolumn{1}{c|}{\textbf{SLSQP}} &
  \multicolumn{1}{c|}{1} &
  50 &
  \multicolumn{1}{c|}{\cellcolor[HTML]{6CC282}46} &
  \multicolumn{1}{c|}{\cellcolor[HTML]{FCF5F8}4} &
  \multicolumn{1}{c|}{\cellcolor[HTML]{6CC282}92\%} &
  \cellcolor[HTML]{FCF5F8}8\% &
  \multicolumn{1}{c|}{\cellcolor[HTML]{15B7F2}45} &
  \multicolumn{1}{c|}{\cellcolor[HTML]{FEF8F8}5} &
  \multicolumn{1}{c|}{\cellcolor[HTML]{15B7F2}90\%} &
  \cellcolor[HTML]{FFF8F9}10\% \\ \midrule
\multicolumn{1}{|c|}{\textbf{}} &
  \multicolumn{1}{c|}{\textbf{COBYLA}} &
  \multicolumn{1}{c|}{2} &
  50 &
  \multicolumn{1}{c|}{\cellcolor[HTML]{74C589}44} &
  \multicolumn{1}{c|}{\cellcolor[HTML]{FCEDF0}6} &
  \multicolumn{1}{c|}{\cellcolor[HTML]{74C589}88\%} &
  \cellcolor[HTML]{FCEDF0}12\% &
  \multicolumn{1}{c|}{\cellcolor[HTML]{44C5F4}38} &
  \multicolumn{1}{c|}{\cellcolor[HTML]{FCE6E8}12} &
  \multicolumn{1}{c|}{\cellcolor[HTML]{44C5F4}76\%} &
  \cellcolor[HTML]{FFE5EA}24\% \\ \midrule
\multicolumn{1}{|c|}{\textbf{}} &
  \multicolumn{1}{c|}{\textbf{L\_BGFS\_B}} &
  \multicolumn{1}{c|}{2} &
  50 &
  \multicolumn{1}{c|}{\cellcolor[HTML]{78C78D}43} &
  \multicolumn{1}{c|}{\cellcolor[HTML]{FCE9EC}7} &
  \multicolumn{1}{c|}{\cellcolor[HTML]{78C78D}86\%} &
  \cellcolor[HTML]{FCE9EC}14\% &
  \multicolumn{1}{c|}{\cellcolor[HTML]{73D4F7}31} &
  \multicolumn{1}{c|}{\cellcolor[HTML]{FAD5D8}19} &
  \multicolumn{1}{c|}{\cellcolor[HTML]{73D4F7}62\%} &
  \cellcolor[HTML]{FED2DC}38\% \\ \midrule
\multicolumn{1}{|c|}{\textbf{}} &
  \multicolumn{1}{c|}{\textbf{SLSQP}} &
  \multicolumn{1}{c|}{2} &
  50 &
  \multicolumn{1}{c|}{\cellcolor[HTML]{7CC890}42} &
  \multicolumn{1}{c|}{\cellcolor[HTML]{FCE5E8}8} &
  \multicolumn{1}{c|}{\cellcolor[HTML]{7CC890}84\%} &
  \cellcolor[HTML]{FCE5E8}16\% &
  \multicolumn{1}{c|}{\cellcolor[HTML]{29BDF3}42} &
  \multicolumn{1}{c|}{\cellcolor[HTML]{FDF0F1}8} &
  \multicolumn{1}{c|}{\cellcolor[HTML]{29BDF3}84\%} &
  \cellcolor[HTML]{FFF0F3}16\% \\ \midrule
\multicolumn{1}{|c|}{\textbf{HO Enc.}} &
  \multicolumn{1}{c|}{\textbf{COBYLA}} &
  \multicolumn{1}{c|}{1} &
  50 &
  \multicolumn{1}{c|}{\cellcolor[HTML]{8CCF9E}38} &
  \multicolumn{1}{c|}{\cellcolor[HTML]{FBD6D9}12} &
  \multicolumn{1}{c|}{\cellcolor[HTML]{8CCF9E}76\%} &
  \cellcolor[HTML]{FBD6D9}24\% &
  \multicolumn{1}{c|}{\cellcolor[HTML]{0EB5F1}46} &
  \multicolumn{1}{c|}{\cellcolor[HTML]{FFFAFB}4} &
  \multicolumn{1}{c|}{\cellcolor[HTML]{0EB5F1}92\%} &
  \cellcolor[HTML]{FFFAFB}8\% \\ \midrule
\multicolumn{1}{|c|}{\textbf{}} &
  \multicolumn{1}{c|}{\textbf{L\_BGFS\_B}} &
  \multicolumn{1}{c|}{1} &
  50 &
  \multicolumn{1}{c|}{\cellcolor[HTML]{B0DDBD}29} &
  \multicolumn{1}{c|}{\cellcolor[HTML]{FAB3B5}21} &
  \multicolumn{1}{c|}{\cellcolor[HTML]{B0DEBE}58\%} &
  \cellcolor[HTML]{FAB3B5}42\% &
  \multicolumn{1}{c|}{\cellcolor[HTML]{5ECEF6}34} &
  \multicolumn{1}{c|}{\cellcolor[HTML]{FBDCDF}16} &
  \multicolumn{1}{c|}{\cellcolor[HTML]{5ECEF6}68\%} &
  \cellcolor[HTML]{FEDAE2}32\% \\ \midrule
\multicolumn{1}{|c|}{\textbf{}} &
  \multicolumn{1}{c|}{\textbf{SLSQP}} &
  \multicolumn{1}{c|}{1} &
  50 &
  \multicolumn{1}{c|}{\cellcolor[HTML]{A8DAB7}31} &
  \multicolumn{1}{c|}{\cellcolor[HTML]{FBBBBD}19} &
  \multicolumn{1}{c|}{\cellcolor[HTML]{A8DAB7}62\%} &
  \cellcolor[HTML]{FBBBBD}38\% &
  \multicolumn{1}{c|}{\cellcolor[HTML]{51C9F5}36} &
  \multicolumn{1}{c|}{\cellcolor[HTML]{FBE1E3}14} &
  \multicolumn{1}{c|}{\cellcolor[HTML]{51C9F5}72\%} &
  \cellcolor[HTML]{FFE0E6}28\% \\ \midrule
\multicolumn{1}{|c|}{\textbf{}} &
  \multicolumn{1}{c|}{\textbf{COBYLA}} &
  \multicolumn{1}{c|}{2} &
  50 &
  \multicolumn{1}{c|}{\cellcolor[HTML]{DCEFE4}18} &
  \multicolumn{1}{c|}{\cellcolor[HTML]{F9888B}32} &
  \multicolumn{1}{c|}{\cellcolor[HTML]{DCEFE4}36\%} &
  \cellcolor[HTML]{F9888B}64\% &
  \multicolumn{1}{c|}{\cellcolor[HTML]{B6E9FB}21} &
  \multicolumn{1}{c|}{\cellcolor[HTML]{F6BCC0}29} &
  \multicolumn{1}{c|}{\cellcolor[HTML]{B6E9FB}42\%} &
  \cellcolor[HTML]{FDB8C7}58\% \\ \midrule
\multicolumn{1}{|c|}{\textbf{}} &
  \multicolumn{1}{c|}{\textbf{L\_BGFS\_B}} &
  \multicolumn{1}{c|}{2} &
  50 &
  \multicolumn{1}{c|}{\cellcolor[HTML]{E0F1E7}17} &
  \multicolumn{1}{c|}{\cellcolor[HTML]{F98587}33} &
  \multicolumn{1}{c|}{\cellcolor[HTML]{E0F1E7}34\%} &
  \cellcolor[HTML]{F98587}66\% &
  \multicolumn{1}{c|}{\cellcolor[HTML]{D1F1FD}17} &
  \multicolumn{1}{c|}{\cellcolor[HTML]{F5B2B7}33} &
  \multicolumn{1}{c|}{\cellcolor[HTML]{D1F1FD}34\%} &
  \cellcolor[HTML]{FDADBE}66\% \\ \midrule
\multicolumn{1}{|c|}{\textbf{}} &
  \multicolumn{1}{c|}{\textbf{SLSQP}} &
  \multicolumn{1}{c|}{2} &
  50 &
  \multicolumn{1}{c|}{\cellcolor[HTML]{DCEFE4}18} &
  \multicolumn{1}{c|}{\cellcolor[HTML]{F9888B}32} &
  \multicolumn{1}{c|}{\cellcolor[HTML]{DCEFE4}36\%} &
  \cellcolor[HTML]{F9888B}64\% &
  \multicolumn{1}{c|}{\cellcolor[HTML]{DEF5FE}15} &
  \multicolumn{1}{c|}{\cellcolor[HTML]{F4ADB2}35} &
  \multicolumn{1}{c|}{\cellcolor[HTML]{DEF5FE}30\%} &
  \cellcolor[HTML]{FDA8BA}70\% \\ \midrule
\multicolumn{1}{|c|}{\textbf{IQP Enc.}} &
  \multicolumn{1}{c|}{\textbf{COBYLA}} &
  \multicolumn{1}{c|}{1} &
  50 &
  \multicolumn{1}{c|}{\cellcolor[HTML]{B8E1C4}27} &
  \multicolumn{1}{c|}{\cellcolor[HTML]{FAABAE}23} &
  \multicolumn{1}{c|}{\cellcolor[HTML]{B8E1C4}54\%} &
  \cellcolor[HTML]{FAABAE}46\% &
  \multicolumn{1}{c|}{\cellcolor[HTML]{AFE7FB}22} &
  \multicolumn{1}{c|}{\cellcolor[HTML]{F7BEC3}28} &
  \multicolumn{1}{c|}{\cellcolor[HTML]{AFE7FB}44\%} &
  \cellcolor[HTML]{FDBAC9}56\% \\ \midrule
\multicolumn{1}{|c|}{\textbf{}} &
  \multicolumn{1}{c|}{\textbf{L\_BGFS\_B}} &
  \multicolumn{1}{c|}{1} &
  50 &
  \multicolumn{1}{c|}{\cellcolor[HTML]{D0EBD9}21} &
  \multicolumn{1}{c|}{\cellcolor[HTML]{FA9496}29} &
  \multicolumn{1}{c|}{\cellcolor[HTML]{D0EBD9}42\%} &
  \cellcolor[HTML]{FA9496}58\% &
  \multicolumn{1}{c|}{\cellcolor[HTML]{CAEFFC}18} &
  \multicolumn{1}{c|}{\cellcolor[HTML]{F5B4B9}32} &
  \multicolumn{1}{c|}{\cellcolor[HTML]{CAEFFC}36\%} &
  \cellcolor[HTML]{FDB0C0}64\% \\ \midrule
\multicolumn{1}{|c|}{\textbf{}} &
  \multicolumn{1}{c|}{\textbf{SLSQP}} &
  \multicolumn{1}{c|}{1} &
  50 &
  \multicolumn{1}{c|}{\cellcolor[HTML]{C0E4CB}25} &
  \multicolumn{1}{c|}{\cellcolor[HTML]{FAA4A6}25} &
  \multicolumn{1}{c|}{\cellcolor[HTML]{C0E4CB}50\%} &
  \cellcolor[HTML]{FAA4A6}50\% &
  \multicolumn{1}{c|}{\cellcolor[HTML]{AFE7FB}22} &
  \multicolumn{1}{c|}{\cellcolor[HTML]{F7BEC3}28} &
  \multicolumn{1}{c|}{\cellcolor[HTML]{AFE7FB}44\%} &
  \cellcolor[HTML]{FDBAC9}56\% \\ \midrule
\multicolumn{1}{|c|}{\textbf{}} &
  \multicolumn{1}{c|}{\textbf{COBYLA}} &
  \multicolumn{1}{c|}{2} &
  50 &
  \multicolumn{1}{c|}{\cellcolor[HTML]{FCFCFF}10} &
  \multicolumn{1}{c|}{\cellcolor[HTML]{F8696B}40} &
  \multicolumn{1}{c|}{\cellcolor[HTML]{FCFCFF}20\%} &
  \cellcolor[HTML]{F8696B}80\% &
  \multicolumn{1}{c|}{\cellcolor[HTML]{FFFFFF}10} &
  \multicolumn{1}{c|}{\cellcolor[HTML]{F2A0A6}40} &
  \multicolumn{1}{c|}{\cellcolor[HTML]{FFFFFF}20\%} &
  \cellcolor[HTML]{FC9AAF}80\% \\ \midrule
\multicolumn{1}{|c|}{\textbf{}} &
  \multicolumn{1}{c|}{\textbf{L\_BGFS\_B}} &
  \multicolumn{1}{c|}{2} &
  50 &
  \multicolumn{1}{c|}{\cellcolor[HTML]{ECF6F2}14} &
  \multicolumn{1}{c|}{\cellcolor[HTML]{F9797B}36} &
  \multicolumn{1}{c|}{\cellcolor[HTML]{ECF6F2}28\%} &
  \cellcolor[HTML]{F9797B}72\% &
  \multicolumn{1}{c|}{\cellcolor[HTML]{EBF9FE}13} &
  \multicolumn{1}{c|}{\cellcolor[HTML]{F4A8AE}37} &
  \multicolumn{1}{c|}{\cellcolor[HTML]{EBF9FE}26\%} &
  \cellcolor[HTML]{FDA2B6}74\% \\ \midrule
\multicolumn{1}{|c|}{\textbf{}} &
  \multicolumn{1}{c|}{\textbf{SLSQP}} &
  \multicolumn{1}{c|}{2} &
  50 &
  \multicolumn{1}{c|}{\cellcolor[HTML]{D4ECDD}20} &
  \multicolumn{1}{c|}{\cellcolor[HTML]{FA9092}30} &
  \multicolumn{1}{c|}{\cellcolor[HTML]{D4ECDD}40\%} &
  \cellcolor[HTML]{FA9092}60\% &
  \multicolumn{1}{c|}{\cellcolor[HTML]{EBF9FE}13} &
  \multicolumn{1}{c|}{\cellcolor[HTML]{F4A8AE}37} &
  \multicolumn{1}{c|}{\cellcolor[HTML]{EBF9FE}26\%} &
  \cellcolor[HTML]{FDA2B6}74\% \\ \bottomrule
\end{tabular}}
\caption{For $6$ and $12$ qubit VRP circuits using SVM on with $2$ layers, the table above shows the  Accuracy and Error with reference to classical minimum (over $50$ iterations) for VQE simulations on variable hamiltonians utilizing Amplitude, Angle, Higher-Order, and IQP encoding schemes, Over the use of COBYLA, SLSQP and L\_BGFS\_B optimizers.}
\label{Accuracy and Error All Encodings Variable hamiltonian}
\end{table*}

\subsection{Inferences from Simulation } \label{Inferences}
As we scan through the results of SVM VRP simulations across the encoding schemes we observe some clear and distinct trends regarding the experiment. The tables \ref{Accuracy and Error All Encodings Fixed Hamiltonian} , \ref{Accuracy and Error All Encodings Variable hamiltonian} summarize the results obtained from the plots of all the encoding schemes used in this experiment. We list these trends as our outcomes of this experiment in the below points

\begin{itemize}
     \item The approach to solving VRP using machine learning is successful and is capable of accomplishing the same or a superior result than the conventional approach using VQE and QAOA.
     
     \item The use of encoding/decoding schemes can serve the purpose of creating superposition and entanglement and eliminate the additional effort required to construct the mixer hamiltonian when solving the VRP using the standard approach of QAOA and VQE.
     
     \item While the standard approach to solving VRP or any combinatorial optimization problem requires a few layers of circuit depth (2 in most cases), we are able to achieve the same on the first layer itself with this approach, proving that it is more efficient than the standard approach.
     
     \item We also observe a distinct trend that as the number of layers increases, the accuracy decreases, which can be used to determine where to limit the optimization depth.
     
     \item Encoding/decoding schemes reduce the number of optimization layers but increase the circuit's complexity by introducing more gates. Therefore, when selecting an encoding scheme, we must take into account the complexity of the generated circuit and the number of required gates, as well as the number of classical resources (memory, CPU) it will require. There must be a trade-off between circuit complexity and the desired problem accuracy.
     
     \item Despite the fact that amplitude encoding provided the greatest accuracy, it could not be used to simulate a $12$-qubit VRP scenario due to the large number of gates required. Angel encoding, on the other hand, was found to be much simpler due to a significantly smaller number of gates, as well as providing excellent accuracy ($96\%$ for COBYLA, and $92\%$ for SLSQP and L\_BGFS\_B in variable hamiltonian simulation) across all the available optimizers. This again demonstrates that the complexity of circuits and the number of gates used are the most important considerations when choosing an encoding/decoding scheme.

     \item It can be noticed that AgE performs the best in terms of circuit complexity and accuracy rates due to the formation of a single layer of superposition. In other encodings (HO, IqpE), we observe multi-layered complex superposition structures, which is the reason for fluctuations or error rates. Also in the fact that increasing layers also increases the superposition structures and therefore decreases the accuracy.
     
     \item Using COBYLA as an optimizer, HO encoding yielded intriguing results with reduced accuracy in circuits with fewer qubits ($6$ qubits) and higher accuracy in circuits with more qubits ($12$ qubits) for both fixed and variable hamiltonian simulations. The trend is disregarded by SLSQP and L\_BGFS\_B. This demonstrates that the algorithm's performance is extremely dependent on the optimizer; therefore, when evaluating the algorithm's performance, the most efficient optimizer should be selected by comparing the available optimizers.
     
     \item The IQP encoding scheme performed the worst in this experiment, with the lowest accuracy and highest error rates among all other encodings used for $1$-layer, $2$-layer, fixed, and variable Hamiltonians simulations. Therefore, the IqpE method cannot be used to solve VRP using QSVM.
     
    \item All of the optimizers used in the experiments performed well across AE, AgE, and HO encodings; however, COBYLA outperformed the other two due to its consistently high level of accuracy, but SLSQP is more resistant to accuracy fluctuations caused by an increase in optimization depth or in the presence of multi-layered circuits.

\end{itemize}

\subsection{Experimental Setup, data gathering, and statistics}
This experiment is conducted within the ambit of the QISKIT framework. while performing the experiment, we used a quantum instance object, and the ansatz runs inside the quantum instance object. A random seed is added to quantum instance to stabilize VQE results.  All the experiments have been run $50+50$ times, one with a fixed Hamiltonian matrix and the other by varying the Hamiltonian matrix. The objective of the experiments is to ensure that the results of experiments are just not dependent on a single Hamiltonian. This is also to ensure that the used circuits achieve classical minimum or near classical minimum regardless of the hamiltonian used. Thus apart from the plots, the tables \ref{Accuracy and Error All Encodings Fixed Hamiltonian}, \ref{Accuracy and Error All Encodings Variable hamiltonian} become the figure of merit. In addition to the many hours of testing and debugging, it is to be noted that the results reported here amounted to $150$ hours of CPU time on a $24$-core AMD workstation using Qiskit's built-in simulators \cite{Qiskit}.

\section{Conclusion}

In this paper, we presented a novel technique for solving VRP through the use of a $6$ and $12$-qubit circuit-based quantum support vector machine (QSVM) with a variational quantum eigensolver for both fixed and variable Hamiltonians. In the experiment, multiple encoding strategies were used to convert the VRP formulation into a QSVM and solve it. In addition, we utilized multiple classical optimizers available within the QISKIT framework to measure the output variation and accuracy rates. Consequently, our machine learning-based approach to resolving VRP has proven fruitful thus far. Using a QSVM to implement a gate-based simulation of a $3$-city or $4$-city VRP on a $6$-qubit or $12$-qubit system accomplishes the goal.  The method not only resolves VRP, but also outperforms the conventional method of resolving VRP via multiple Optimization phases involving only VQE and QAOA. In addition, selecting appropriate encoding methods establishes the optimal balance between circuit complexity and optimization depth, thereby enabling multiple approaches to solve CO problems using machine learning techniques.

\section*{Acknowledgment}
The authors are grateful to the IBM Quantum Experience platform and their team for developing the Qiskit platform and providing open access to their simulators for running quantum circuits and performing the experiments reported here \cite{Qiskit}.

\bibliographystyle{IEEEtran}
\bibliography{IEEE}
\end{document}